\documentclass[a4paper, 12pt]{article}

\usepackage{xcolor}
\usepackage{authblk}
\usepackage{graphicx}
\usepackage{epstopdf}
\usepackage{amsmath}
\usepackage{amssymb}
\usepackage{amsfonts}
\usepackage{titlesec}
\usepackage{changepage}
\usepackage[square,sort,comma,numbers]{natbib}
\usepackage[hidelinks]{hyperref}
\usepackage[font=small]{caption}

\titleformat{\subsection}[runin]{\bfseries}{}{}{}[.]

\let\citeleft=(
\let\citeright=)

\newcommand{\norm}[1]{\left\lVert#1\right\rVert}                                                

\newcommand{\tab}{\hspace{1cm}}

\DeclareMathOperator*{\argmin}{arg\,min}                                                        

\newcommand{\fig}[4]{
  \begin{figure}[ht!] \centering
  \includegraphics[width=#2\textwidth]{#1}
  \caption{
    #4
  }
  \label{#3}
  \end{figure}
}

\newcommand{\customsubscript}[2]{$\text{#1}_{#2}$}

\AppendGraphicsExtensions{.tif}

\begin{document}

\pdfinfo{ 
  /Author (Siddharth Iyer, Daniel Polak, Congyu Liao, Jonathan I. Tamir, Stephen F. Cauley, Borjan Gagoski, Wei-Ching Lo, Berkin Bilgic and Kawin Setsompop)
  /Title (Wave-encoding and Shuffling Enables Rapid Time Resolved Structural Imaging)
}

\title{\vspace{-2cm}\Large Wave-encoding and Shuffling Enables Rapid Time Resolved Structural Imaging}
\date{}

\author[1, 2, 7, 8]{\normalsize Siddharth Iyer}
\author[1, 3]{\normalsize Daniel Polak}
\author[1, 7, 8]{\normalsize Congyu Liao}
\author[4]{\normalsize Jonathan I. Tamir}
\author[1]{\normalsize Stephen F. Cauley}
\author[5]{\normalsize Borjan Gagoski}
\author[6]{\normalsize Wei-Ching Lo}
\author[1]{\normalsize Berkin Bilgic}
\author[1, 7, 8]{\normalsize Kawin Setsompop}

\affil[1] { \footnotesize
  Athinoula A. Martinos Center for Biomedical Imaging, Charlestown, MA, United States
}

\affil[2] { \footnotesize
  Department of Electrical Engineering and Computer Science, Massachusetts Institute of Technology,
  Cambridge, MA, United States
}

\affil[3] { \footnotesize
  Siemens Healthcare GmbH, Karl-Schall-Straße 6, 91052 Erlangen, Germany
}

\affil[4] { \footnotesize
  Electrical and Computer Engineering, University of Texas at Austin, 2501 Speedway,
  Austin, TX, United States
}

\affil[5] { \footnotesize
  Boston Children's Hospital, Boston, Massachusetts, Boston, MA, United States
}

\affil[6] { \footnotesize
  Siemens Medical Solutions, Boston, MA, United States
}

\affil[7] { \footnotesize
  Department of Radiology, Stanford University,
  Stanford, CA, United States
}

\affil[8] { \footnotesize
  Department of Electrical Engineering, Stanford University,
  Stanford, CA, United States
}

\maketitle

\vfill\vspace{-1.5cm}

\noindent
{\footnotesize \textit{Running head:} Wave-encoding and Shuffling Enables Rapid Time Resolved Structural Imaging}

\noindent
{\footnotesize \textit{Address correspondence to:}        \\
Siddharth Iyer                                            \\
Department of Electrical Engineering and Computer Science \\
Massachusetts Institute of Technology                     \\
Cambridge, MA 02129 USA                                   \\
\href{mailto:ssi@mit.edu}{ssi@mit.edu}}

\noindent
{\footnotesize This work was supported in part by NIH research grants: R01MH116173, R01EB020613, R01EB019437, U01EB025162, P41EB030006, and the shared
instrumentation grants: S10RR023401, S10RR019307, S10RR019254, S10RR023043.}

\vfill\vspace{-0.5cm}

\noindent
{\footnotesize Approximate word count: 200 (Abstract) 6090 (Body)}

\vfill\vspace{-0.5cm}

\noindent
{\footnotesize Parts of this work have been presented at the ISMRM Annual Conferences of 2018 and 2019.}

\vfill\vspace{-0.5cm}

\noindent
Submitted to \textit{Magnetic Resonance in Medicine} as a Full Paper.

\newpage

\section{Abstract}
\subsection{Purpose}
$T_2$-Shuffling reconstructs multiple $T_2$-weighted images from a single volumetric fast spin-echo
(3D-FSE) scan.
Wave-CAIPI achieves good reconstruction at high accelerations by better utlizing coil-sensitivity
information using additional sinusoidal gradients.
In this work, the $T_2$-Shuffling model is augmented with wave-encoding to achieve higher
acceleration capability.
The resulting ``Wave-Shuffling'' approach is applied to 3D-FSE and Magnetization-Prepared Rapid
Gradient-Echo (MPRAGE) to achieve rapid, 1 mm-isotropic resolution, time-resolved structural
imaging.
\vspace{-1cm}
\subsection{Theory and Methods}
Wave-encoding is incorporated into Shuffling and the wave-encoding parameters are optimized
to achieve higher acceleration for the target time-resolved imaging application.
3D-FSE and MPRAGE sequences are modified to enable sinusoidal wave-gradients and random
sub-sampling of k-space.
In-vivo Shuffling and Wave-Shuffling under-sampling MPRAGE experiments are performed to
demonstrate the better conditioning provided by wave-encoding.
In-vivo Wave-Shuffling 3D-FSE data are also acquired and reconstructed.
A wave-encoding calibration method to account for system hardware imperfections suitable
for random under-sampling acquisitions without a densely-sampled, low-resolution calibration
region is proposed.
\vspace{-1cm}
\subsection{Results}
For brain imaging application with 32-channel coil at 3T, Wave-Shuffling MPRAGE provides
comparable reconstruction at $\approx8\times$ higher acceleration compared to Shuffling.
Wave-Shuffling is successfully applied to MPRAGE and 3D-FSE to recover high-quality
time-series of images.
\vspace{-1cm}
\subsection{Conclusion}
Wave-Shuffling enables rapid, time-resolved structural imaging.
\vspace{-1cm}
\subsection{Keywords}
{\it Wave-CAIPI, $T_2$-Shuffling, MPRAGE, 3D-FSE, Time-resolved structural imaging}
\clearpage

\section{Introduction\label{sec:intro}} 
A number of widely used MR sequences such as Fast Spin Echo (FSE) \cite{mugler2014optimized}
and Magnetization-Prepared Rapid Gradient-Echo (MPRAGE) \cite{muglermp} reduce scan time by
performing k-space encoding along a multi-echo train.
For high resolution acquisitions, signal variations along the long echo trains can cause
significant image blurring.
To solve this, spatio-temporal model-based techniques to improve image sharpness while also
recovering additional multiple image contrasts across the echo train have emerged.

\tab
An inherent challenge in high resolution spatio-temporal imaging is acquiring enough k-space
samples for each echo to resolve both spatial and temporal dimensions in a clinically
feasible time frame.
To overcome this, subspace methods have been introduced to greatly reduce the
dimensionality of the temporal domain during reconstruction \cite{4193454}.
Despite the smaller dimensionality, the problem is still often ill-posed.
Consequently, subspace methods have been used with random sampling and sparsity constrained
optimization for better time-resolved imaging and accelerated  $T_1$ and $T_2$ parameter
mapping \cite{zhao2015accelerated}.

\tab
Note that in the above context and for the rest of the work, ``time-resolved'' refers to
reconstructing individual echo-images from a single acquisition as the signal varies over
the echo-train.
In other words, the tissue dynamics over the multi-echo train are being resolved at the
various echo-times over the multi-echo train.
This is not related to real-time acquisitions.

\tab
$T_2$-shuffling is a promising subspace method that has been applied to clinical volumetric
FSE (3D-FSE) imaging \cite{tamir2017t2, tamir2019}.
By leveraging random under-sampling in the phase-encode and partition directions, and applying
subspace reconstruction with the locally low-rank constraint, the method time-resolves an
image-series that captures the signal evolution across the echo train and avoids image blurring
artifacts.
By combating image blurring, $T_2$-Shuffling has enabled clinical use of 3D-FSE for
pediatric knee MRI by producing diagnostic images in $6-7$ minutes with similar quality to
conventional 2D-FSE acquisition that requires $13$ minutes of scan time \cite{bao2017fast}.
The Shuffling approach from $T_2$-Shuffling has also been applied to the MPRAGE sequence to extract
a time-series of $T_1$-weighted images \cite{cao2018shuffled}.

\tab
In addition to the use of subspace modeling, new sequence-based approaches have also been proposed
to enable fast time-resolved imaging.
MPnRAGE \cite{kecskemeti2016mpnrage} is a multi-echo MPRAGE-like acquisition that leverages radial
sampling and a sliding window reconstruction to recover the image time-series across the echo train.
Echo Planar Time Resolved Imaging (EPTI) \cite{wang2019echo} improves on conventional multi-shot
Echo Planar Imaging by optimizing k-space and time sampling to exploit temporal
correlations during reconstruction to recover distortion free, multi-contrast time series of
EPI images.
Further more, subspace modeling has been incorporated into EPTI to achieve higher rates of
acceleration \cite{dong2020echo}.

\tab
In the pursuit of rapid, high resolution spatio-temporal imaging, advances in parallel-imaging
that reduce the number of k-space samples needed to spatially resolve the underlying image in
non time-resolved acquisitions can be leveraged to improve the encoding of accelerated
time-resolved imaging.
Wave-CAIPI \cite{bilgic2015wave} is a parallel-imaging technique that utilizes additional
sinusoidal gradients during the readout to spread aliasing in the readout direction.
This is shown to improve the acceleration capabilities of parallel-imaging through better
utility of 3D coil-sensitivity information, resulting in a better posed inverse problem with
excellent g-factor performance.
Wave-CAIPI has been successfully applied to volumetric 3D-FSE to achieve nine-fold
acceleration at $3T$ \cite{polak2019highly}, and to MPRAGE to enable nine-fold and twelve-fold
acceleration at $3T$ and $7T$ respectively \cite{polak2019highly, polak2018wave}.

\tab
However, conventional Wave-CAIPI acquisitions cannot fully utilize the better conditioning
provided by wave-encoding.
Simulation studies have shown that g-factor performance is largely determined by the maximum
gradient amplitude ($G_{\max}$) of the sinusoidal wave-gradient and is independent of the
number of sinusoidal cycles \cite{polak2019highly}.
To achieve the maximum possible $G_{\max}$, lower wave cycle numbers are required due to
gradient hardware slew rate limitations.
When utilizing such encoding, conventional Wave-CAIPI acquisitions, where the data are
acquired across a long echo train, suffer from undesirable image ringing artifacts due to
the interaction between signal modulation along the echo train from $T_1/T_2$ signal
relaxation and the voxel spreading effect from wave-encoding \cite{polak2019highly}.
This ``Signal-Mixing'' artifact becomes significant when a high $G_{\max}$ and a low cycle
number wave-encoding  is used, limiting the amount of encoding performance that can be
practically achieved.
However, this can be overcome by time-resolving the echo-train during reconstruction.
In particular, while Wave-CAIPI was developed as a natural extension to controlled aliasing
methods like Bunch Phase Encoding \cite{moriguchi2006bunched} and 2D-CAIPIRINHA
\cite{breuer2005controlled, breuer2006controlled}, the sinusoidal gradients have also been
successfully applied to data acquired with random k-space sampling to achieve high rates of
acceleration with good reconstruction \cite{bilgic2016optimized, chen2018self}.
This suggests that wave-encoding can be integrated into spatio-temporal model-based
techniques to be robust to Signal-Mixing artifacts while improving the acceleration
performance of time-resolved applications.

\tab
When utilizing wave-encoding, hardware imperfections can cause errors in the sinusoidal
gradient resulting in detrimental image ringing artifacts if not accounted for in the
reconstruction.
Auto-calibrated PSF (or ``AutoPSF'') methods \cite{cauley2017autocalibrated, chen2018self} have
been proposed to estimate these gradient imperfections directly from the under-sampled data,
but these approaches cannot be directly applied to acquisitions with random under-sampling
without a fully sampled calibration region due to high computational cost.
Consequently, to avoid the need to acquire additional calibration data, a computationally
efficient wave-calibration method for the mentioned conditions is required.

\tab
In order to fully utilize the better conditioning provided by high $G_{\max}$ wave-encoding,
``Wave-Shuffling'' is proposed where wave-encoding is incorporated into the Shuffling model
with the temporal subspace and random sub-sampling.
By temporally-resolving the underlying signal during the reconstruction, ``Wave-Shuffling''
is shown to be robust to the ``Signal Mixing'' artifacts, enabling higher $G_{\max}$ compared
to conventional Wave-CAIPI.
The $G_{\max}$ of wave-encoding is then optimized to achieve significantly higher acceleration
capability compared to standard Shuffling while retaining the multi-contrast and clinically
desirable blur-free reconstruction provided by Shuffling \cite{bao2017fast}.
To enable Wave-Shuffling application that is robust to hardware imperfections, a computationally
efficient modification of the AutoPSF approach by Cauley et al.\ \cite{cauley2017autocalibrated}
is proposed.

\section{Theory}~\\
A brief review of the wave-encoding and the Shuffling model is provided, along with a description
on how these approaches are combined to form Wave-Shuffling.  
\subsection{Wave-encoding\label{sec:wavetheory}}~\\ 
In wave-encoding, additional sinusoidal $(G_y, G_z)$ gradients are applied during the readout as
depicted in Figure \ref{fig:explainer}(A).
This induces a voxel spreading effect along the readout direction and the amount of spreading
increases linearly as a function of the spatial $(y, z)$ locations.
A two-dimensional $(x, y)$ example of this is shown in Figure \ref{fig:explainer}(B) for $G_y$
wave-encoding.
In an accelerated acquisition, this voxel spreading effect acts to increase the distance between
the aliased voxels to allow for better use of coil sensitivity information for improved
reconstruction \cite{bilgic2015wave}.

\tab
The voxel spreading can be modeled by a simple point-wise multiplication with a wave point-spread
function (Wave-PSF) \cite{bilgic2015wave}.
This Wave-PSF, denoted $W$, is described in the hybrid $(k_x, y, z)$ space and consists of
phase-only entries with unitary complex magnitude.

\tab
Let $m$ be the underlying image, $v$ be the acquired data, $S$ denote coil-sensitivity maps,
$F_x$ denote the forward Fourier transform along the readout
direction, $F_{yz}$ denote the forward Fourier transform along the phase encode and partition
directions, and $M$ be the k-space sampling mask.
Let $R$ be the resizing operator that resizes in the readout dimension to a larger field of
view (FOV) to account for the larger $FOV_x$ required to capture the spreading effect of
wave-encoding.
The wave-encoding forward model is then,
\begin{equation}
v = MF_{yz}WF_xRSm
\label{eq:wave}
\end{equation}
\tab
When using low $G_{\max}$ wave-encoding, it suffices to only resize the readout dimension
to a larger field of view (FOV) to account the spreading effect of wave-encoding.
In the regime of high $G_{\max}$ wave-encoding, in addition to zero-padding the readout
dimension in image space $(x)$, it is necessary to zero-pad the phase-encode $(k_y)$ and partition
encode $(k_z)$ directions in k-space to ensure that the multiplication with the Wave-PSF $(W)$,
which is in  $(k_x, y, z)$ space, correctly performs linear convolution in $(x, k_y, k_z)$ space.
This zero-padding takes into account the k-space wrap around effect in $k_y$ and $k_z$ of
large $G_{\max}$ wave-encoding in combination with the k-space interpolation performed by
parallel-imaging.
\subsection{Shuffling}~\\ 
Shuffling models the magnetization temporal dynamics of the underlying image as it evolves over
an echo train.
The signal evolutions of tissues are observed to be well represented by a low-dimensional subspace
for which an orthonormal basis can be calculated using the Singular Value Decomposition (SVD)
\cite{tamir2017t2}.
This is depicted in Figure \ref{fig:explainer}(C).
The methodology this work used to pick temporal subspace rank is discussed in the \nameref{sec:subdesign}
sub-section.
This orthonormal temporal basis, denoted $\Phi$, is used in the Shuffling forward model as
follows:
\begin{equation}
v = MF_{xyz}S\Phi\alpha
\label{eq:shfl}
\end{equation}
Here, $\alpha$ are the ``coefficient'' images that are passed through $\Phi$ to recover the
time-series of images over the echo train, $M$ denotes the spatio-temporal sampling mask,
and $v$ denotes the acquired multi-echo k-space data.

\tab
Even with the incorporation of the temporal basis $(\Phi)$, Equation \eqref{eq:shfl} is still
ill-posed for applications that can be performed in a reasonable time frame with a limited
number of k-space encoding lines (similar to that of conventional non time-resolved
acquisitions).
To overcome this, random sampling along both k-space and the echo-time dimensions is employed
with a regularized reconstruction and a sparsity promoting transform $(\Psi)$ such as locally
low-rank (LLR) to better recover the underlying coefficient images $(\alpha)$.
\begin{equation}
\min\limits_\alpha \norm{v - MF_{xyz}S\Phi\alpha}_2^2 + \lambda \norm{\Psi \alpha}_1
\label{eq:shflrecon}
\end{equation}
\tab
The expected performance of Equation \eqref{eq:shflrecon} for a given Shuffling acquisition can
be qualitatively analyzed by studying its Point Spread Function (PSF), which can be calculated
as follows \cite{tamir2017t2}:
\begin{equation}
\text{PSF}_n = (\Phi^* F_{xyz}^*MF_{xyz}\Phi) \; \delta_n
\label{eq:shfltpsf}
\end{equation}
\tab
The delta image $(\delta_n)$ is zero at all entries except at the center image of the $n^{th}$
coefficient, where it is one.
The ideal response would be the same delta image for the $n^{th}$ coefficient with zero cross talk
across the other coefficient images.
However, due to acceleration and random sampling, the resulting PSF will have noise-like incoherent
side-lobes in the $n^{th}$ coefficient image with signal aliasing to the other coefficients.
The \textbf{peak-to-side-lobe ratio} of \customsubscript{PSF}{n} refers to the largest absolute voxel
value over all coefficients of \customsubscript{PSF}{n} not including the voxel where $\delta_n$ is
non-zero.
The lower the peak-to-side-lobe of the PSF over all values of $n$, the weaker the noise-like
incoherent aliasing artifacts and the more likely an $l_1-$regularized reconstructions like
Equation \eqref{eq:shflrecon} will be able to ``de-noise'' the incoherent aliasing artifacts to
recover the true underlying images.
\subsection{Wave-Shuffling}~\\ 
In Wave-Shuffling, the Shuffling forward model (Equation \eqref{eq:shfl}) is augmented with the
sinusoidal gradients of Wave-CAIPI (Equation \eqref{eq:wave}).
Akin to Shuffling, Wave-Shuffling is solved using a regularized reconstruction and a
sparsifying transform $(\Psi)$.
\begin{equation}
\min\limits_\alpha \norm{v - MF_{yz}WF_xRS\Phi\alpha}_2^2 +
\lambda \norm{\Psi \alpha}_1
\label{eq:wshflrecon}
\end{equation}
\tab
The conditioning benefits of the wave sinusoidal gradients should allow for better utilization of
the coil sensitivity profiles, allowing for higher under-sampling and hence faster acquisition times
in comparison to standard Shuffling.

\tab
The PSF for Wave-Shuffling can be calculated as follows.
\begin{equation}
\text{WPSF}_n = (\Phi^* R^* F_x^* W^* F_{yz}^*MF_{yz}W F_{x} R \Phi) \; \delta_n
\label{eq:wshfltpsf}
\end{equation}
\subsection{Wave-Encoding Parameter Optimization\label{sec:waveparamopt}}~\\ 
As discussed in the \nameref{sec:intro} section, better g-factor performance is
achieved using sinusoidal wave-gradients with a higher $G_{\max}$ which require lower
wave cycle numbers due to gradient hardware slew rate limitations.
When utilizing such encoding, conventional Wave-CAIPI acquisitions suffer from ringing
artifacts, denoted ``Signal-Mixing'', due to the interaction between the $T_1/T_2$ signal
relaxation over the echo-train and the voxel spreading effect from wave-encoding
\cite{polak2019highly}.
Wave-Shuffling mitigates this artifact as the data are time-resolved across the echo-train.
Consequently, more efficient wave-encoding can be achieved by optimizing the wave-gradient
maximum amplitude and the number of sinusoidal cycles.
The peak-to-side-lobe ratio of the PSF of Wave-Shuffling (Equation \eqref{eq:wshfltpsf})
as a function of wave-encoding parameters is used as a metric for parameter optimization.


\subsection{Wave-PSF Calibration\label{sec:autopsf}}~\\ 
Hardware imperfections can cause errors in the sinusoidal gradients of wave-encoding, resulting
in image ringing artifacts if not accounted for in the reconstruction.
Auto-calibrated PSF (or ``AutoPSF'') methods \cite{cauley2017autocalibrated, chen2018self} have
been proposed to estimate these gradient imperfections directly from the under-sampled wave data.
However, these approaches cannot be directly applied to Wave-Shuffling as data are acquired with
random under-sampling without a fully sampled calibration region.
To enable Wave-Shuffling application that is robust to hardware imperfections, an efficient
AutoPSF method is developed for Wave-Shuffling by modifying the approach proposed by Cauley et al.
in \cite{cauley2017autocalibrated}.
Below, a brief review of the AutoPSF approach in \cite{cauley2017autocalibrated} is provided along
with the proposed extension to enable it to work efficiently with Wave-Shuffling.

\tab
The Wave-PSF $(W)$, used in Equation \eqref{eq:wave}, is constructed in the $(k_x, y, z)$ domain
\cite{bilgic2015wave} as follows:
\begin{equation}\begin{array}{rcl}
W(k_x(t), y, z) &=& \exp
\left\{1i \times 2\pi \times \left[ P_y(k_x(t)) \cdot y + P_z(k_x(t)) \cdot z\right]\right\} \\
P_y(t) &=& \frac{\gamma}{2\pi} \int_0^t G_y(\tau) d\tau \\
P_z(t) &=& \frac{\gamma}{2\pi} \int_0^t G_z(\tau) d\tau
\label{eq:wavegen}
\end{array}\end{equation}
Here, $G_y$ and $G_z$ are the sinusoidal gradients depicted in Figure \ref{fig:explainer}(A),
$\gamma$ is the Larmor frequency and $k_x(t)$ denotes the $(k_x)$ point sampled at
readout-time $(t)$.
Cauley et al.\ proposed a joint optimization technique that jointly solves for the underlying
image and the imperfections of the sinusoidal gradients during reconstruction
\cite{cauley2017autocalibrated}.
The key innovation of the approach is the observation that a select small set of Fourier
coefficients accurately model the sinusoidal wave-gradients along with the related gradient
hardware errors.
Let $c_y$ and $c_z$ denote the select small set of Fourier coefficients of the $G_y$ and $G_z$
gradients respectively, and let $F$ denote the Fourier transform.
The representation of the gradient wave-forms is as follows:
\begin{equation}\begin{array}{rcl}
G_y(t) &=& \left[F(c_y)\right](t) \\
G_z(t) &=& \left[F(c_z)\right](t)
\end{array}\end{equation}
\tab
The derived $(G_y, G_z)$ gradient wave-forms are used to generate the Wave-PSF $(W)$ as in Equation
\eqref{eq:wavegen}.
Let $W(c)$ denote the Wave-PSF derived from gradient wave-forms obtained from $c = (c_y, c_z)$.
AutoPSF iterates between minimizing the underlying image $(m)$ and the Fourier coefficients $(c)$ to
reduce the data consistency error of the reconstruction.
Using the wave forward model in Equation \eqref{eq:wave}, this alternating minimization can
be expressed as,
\begin{equation}\begin{array}{rrclr}
\text{Data Consistency:} & m&=&\argmin_m&\norm{v - MF_{yz}W(c)F_xRSm} \\
\text{Fourier Update:}   & c&=&\argmin_c&\norm{v - MF_{yz}W(c)F_xRSm}
\label{eq:autopsf}\end{array}\end{equation}
\tab
The first image $(m)$ is reconstructed with Fourier estimates $(c)$ that are derived from assuming
ideal gradients $(G_y, G_z)$ with no gradient hardware errors.
This image $(m)$ is then used to update the coefficients $(c)$ in the Fourier Update step, and
the new coefficients are used to update the image $(m)$ in the Data Consistency step and so on.
This alternating minimization approach is shown to provide high-quality reconstruction for the
conventional Wave-CAIPI reconstruction \cite{cauley2017autocalibrated}.

\tab
In the Data Consistency step of Equation \eqref{eq:autopsf}, instead of solving the entire imaging
volume at once, AutoPSF leverages the structured aliasing of Wave-CAIPI for fast computation by
selecting a small set of representative voxels from which an estimate of the data-consistency error
is calculated.
This does not extend to Wave-Shuffling as the use of random sampling results in incoherent aliasing
artifacts (as opposed to the structured aliasing in Wave-CAIPI).
Consequently, it is not possible to select a small subset of representative voxels and a full volume
Wave-Shuffling reconstruction (Equation \eqref{eq:wshflrecon}) in lieu of the Data Consistency in
Equation \eqref{eq:autopsf} is required to get an accurate representation of the data-consistency
error.
This dramatically increases the total computational cost of AutoPSF.
To overcome this computational difficulty, the following PSF-calibration technique is proposed with
the goal of robust and data-driven wave calibration for Wave-Shuffling acquisitions.


\tab
Let $c_0$ be the Fourier coefficients of the ideal wave sinusoidal gradients (assuming no hardware
errors) and let $c_*$ be the Fourier coefficients of the corrected wave-gradients obtained from the
optimization presented in Equation \eqref{eq:autopsf}.
Let $c_e$ denote the difference between $c_0$ and $c_*$.
\begin{equation}
c_e = (c_* - c_0).
\label{eq:wavedif}
\end{equation}
\tab
Let $W_*, W_0$ and $W_e$ denote the Wave-PSFs derived from $c_*, c_0$ and $c_e$ respectively.
Utilizing the linearity of the Fourier transform, Equation \eqref{eq:wavedif} in conjunction with
Equation \eqref{eq:wavegen} implies the following:
\begin{equation}\begin{array}{rcl}
W_*(k_x(t), y, z)
  &=& \exp \left\{1i \times 2\pi \times \left( \left[F((c_*)_y)\right](k_x(t)) \cdot y + \left[F((c_*)_z)\right](k_x(t)) \cdot z\right)\right\} \\
  &=& \exp \left\{1i \times 2\pi \times \left( \left[F((c_e + c_0)_y)\right](k_x(t)) \cdot y + \left[F((c_e + c_0)_z)\right](k_x(t)) \cdot z\right)\right\} \\
  &=& \exp \left\{1i \times 2\pi \times \left( \left[F((c_0)_y)\right](k_x(t)) \cdot y + \left[F((c_0)_z)\right](k_x(t)) \cdot z\right)\right\} \times \\
  & & \exp \left\{1i \times 2\pi \times \left( \left[F((c_e)_y)\right](k_x(t)) \cdot y + \left[F((c_e)_z)\right](k_x(t)) \cdot z\right)\right\}        \\
  &=& W_0 W_e 
\end{array}\label{eq:psfrec}\end{equation}
\tab
This suggests that the artifacts caused by an incorrect Wave-PSF can be modeled by a convolution
with $W_e$, and that it may be possible to estimate a deconvolution to recover $W_*$ using
Equation \eqref{eq:psfrec}.
Doing so re-casts the problem of wave calibration to a deconvolution problem.
Let $(\alpha_0)$ be the initial Wave-Shuffling reconstruction performed using $(W_0)$.
This reconstruction is expected to have ringing artifacts due to system hardware imperfections.
It is assumed that these artifacts are well approximated by a convolution with the Wave-PSF
$(W_e)$, and finding the inverse PSF that mitigates these ringing artifacts will allow for the
recovery of $(W_e)$.
In particular, since $(W_e)$ is a phase only PSF, the inverse PSF is simply the element wise complex
conjugate PSF $(W_e^*)$.
Consequently, finding a PSF that mitigates the ringing artifacts of the initial reconstruction
$(\alpha_0)$ is expected to be a good approximation of $(W_e^*)$ and consequently $(W_e)$.

\tab
In formulating an optimization procedure that can effectively find the Wave-PSF $(W_e^*)$ that
mitigates the ringing artifacts of $(\alpha_0)$, note that these artifacts will be exaggerated by
applying an edge-detector along the readout direction.
Let this edge detector be $T$.
Consider the following optimization problem:
\begin{equation}
c_e = \argmin_c \norm{T\left(F_x^{-1} W_e^*(c) F_x \alpha_0\right)}_1
\label{eq:waveopt}
\end{equation}
\tab
This optimization attempts to correct for Wave-PSF errors by performing wave deconvolution as a
post-processing step (by applying $F_x^{-1} W_e^*(c) F_x$).
After deconvolving $(\alpha_0)$ with the estimated PSF $W_e^*(c)$, the edge detector is applied to
amplify the remaining artifacts.
Since there are only a limited number of coefficients $(c)$, Equation \eqref{eq:waveopt} is efficiently
solved with the Nelder-Mead simplex algorithm.
The optimization is expected to recover the desired difference coefficients $(c_e)$ as other
values of the coefficients $(c)$ in the optimization (Equation \eqref{eq:waveopt}) are expected to
induce additional undesired voxel spreading.
The corrected Wave-PSF $(W_*)$ can then be recovered as in Equation
\eqref{eq:psfrec} to be used in a Wave-Shuffling reconstruction.

\tab
In this work, the following edge detector is used.
Let $D_x$ denote the finite difference along the readout direction and let $\mu$ be a positive scalar.
The edge detector $T$ is defined as follows:
\begin{equation}
T(m) = \text{sigmoid}\left(\mu \times \left|D_x(m)\right|\right) - \frac{1}{2}
\label{eq:edgedetector}
\end{equation}
A $\mu$ value of $100$ is empirically found to sufficiently amplify wave-related ringing artifacts
while not being over-sensitive to noise.

\tab
For additional computational efficiency, the separability of wave-encoding is utilized.
The $(G_y, G_z)$ sinusoidal gradients induce voxel spreading independently of each other.
Consequently, the Wave-PSF $(W)$ can be split into separate $(k_x, y)$ and $(k_x, z)$ PSFs
as follows.
\begin{equation}
W(k_x, y, z) = W_y(k_x, y) W_z(k_x, z)
\end{equation}
\tab
Therefore, to calibrate for $W_y$ without being affected by $W_z$, the optimization in
Equation \eqref{eq:waveopt} is performed on a single slice of the three-dimensional volume
at $z = 0$ spatial position.
The similar process is repeated for $W_z$ by extracting the slice at $y = 0$ spatial position.

\tab
For succinctness, this method of Wave-Calibration will be referred to as ``ShflPSF''.

\section{Methods\label{sec:methods}}
\subsection{Subspace Design\label{sec:subdesign}}~\\ 
It is helpful to first broadly describe the methodology used to generate the low-rank
temporal basis (the $\Phi$ in Equations \eqref{eq:shfl}, \eqref{eq:shflrecon} and
\eqref{eq:wshflrecon}) for the acquisitions of interest.
\subsubsection{\small \it MPRAGE}
For all the following MPRAGE experiments, the Bloch equation was used to simulate the signal
evolution across the echo train during the inversion recovery of MPRAGE using a flip angle of
$9^\circ$ and $T_1$ relaxation values uniformly sampled across the range $[50, 5000]$
milliseconds at 10 millisecond intervals.
These are the expected $T_1$ values of tissues in the brain, and are similar to those used in
prior work \cite{ye2016accelerating}.
In order to be robust to $B_1^+$ variation, the signal simulations were repeated for MPRAGE flip
angles across the range $[6.3, 11.7]$ degrees with a step size of $0.9^\circ$.
Unless otherwise specified, an echo spacing of $8.1$ ms was assumed for the simulations.

\tab
The Singular Value Decomposition (SVD) of the simulated signal ensemble was taken and the singular
vectors were used to derive the low-rank temporal subspace.
The signal ensemble was projected onto the subspace spanned by the first $k$ singular vectors
corresponding to the $k^{th}$ largest singular values and the normalized root mean squared
error (NRMSE) for each signal in the ensemble was calculated.
The maximum NRMSE over all signals was used as a metric to determine the number of basis vectors to use.
\subsubsection{\small \it 3D-FSE}
For the following 3D-FSE experiment, note that a variable flip angle train was utilized
to extend the signal duration over the echo-train \cite{mugler2014optimized}.
Consequently, it is possible to recover signal at echo-times much longer than conventional
spin-echo imaging.
The signal evolutions were simulated using Extended Phase Graphs \cite{weigel2015extended}.
The flip angle train used during acquisition was saved and the simulations were generated
assuming $T_1$ values of $[200, 400, 800, 1400, 2000, 4000]$ in milliseconds, and $T_2$ values in
the range $[20, 2000]$ milliseconds with a step size of $2$ milliseconds.
Similarly to the MPRAGE case, these are the expected $T_1$ and $T_2$ values of tissues in the brain,
and are similar to those used in prior work \cite{ye2016accelerating}.
(FSE has a weak dependence on $T_1$ \cite{tamir2017t2}, hence a smaller range of $T_1$ values were
used for simulation.)
The first 45 echoes (of 256) were discarded as they significantly ill-conditioned the reconstruction.
The Singular Value Decomposition (SVD) of the simulated signal ensemble was taken and the
singular vectors were used to derive the low-rank temporal subspace.
The signal ensemble was projected onto the subspace spanned by the first $k$ singular vectors
corresponding to the $k^{th}$ largest singular values and the NRMSE for each signal in the ensemble
was calculated.
The maximum NRMSE over all signals was used as a metric to determine the number of basis vectors to use.
\subsection{PSF Analysis and Wave-Encoding Parameter Optimization\label{sec:tpsfmethod}}~\\ 
To examine the benefit of incorporating wave-encoding into the Shuffling model, the PSF of MPRAGE
Shuffling and MPRAGE Wave-Shuffling was calculated for a k-space sampling mask where $25\%$ of all
phase and partition encodes was sampled uniformly at random and each encode was sampled exactly once.
The matrix size assumed was $256 \times 256 \times 256$ with a turbo factor (TF) of $256$.
The PSFs calculations were implemented using MATLAB (MathWorks, Natick, MA).
For simplicity and to aid visualization, only the first two basis vectors (corresponding to the
two largest singular values) are used.

\tab
Delta images $\delta_1$ and $\delta_2$ were passed through the Shuffling and Wave-Shuffling forward
models (Equations \eqref{eq:shfltpsf} and \eqref{eq:wshfltpsf}) to obtain ``\customsubscript{PSF}{1}''
and ``\customsubscript{PSF}{2}'' for Shuffling, and ``\customsubscript{WPSF}{1}'' and
``\customsubscript{WPSF}{2}'' for Wave-Shuffling respectively.

\tab
The peak-to-side-lobe ratios of the PSFs were calculated and the result of the following Equation
\eqref{eq:tpsfcost} as a function of wave cycle number and maximum wave amplitude are visualized.
\begin{equation}
\max_n \left[\text{peak-to-side-lobe}\left(\text{WPSF}_n\right)\right]
\label{eq:tpsfcost}\end{equation}
This value was used to optimize the parameters for wave-encoding, as discussed in the
\nameref{sec:waveparamopt} sub-section.
For a given cycle number, the maximum amplitude was determined by either gradient hardware slew
rate limitations or the peripheral nerve stimulation limit.
\subsection{In-Vivo Experiments}~\\ 
The 3D-FSE and MPRAGE sequences were modified to enable sinusoidal wave-gradients and random
sub-sampling of k-space.
Three-dimensional Standard Wave MPRAGE, MPRAGE Shuffling, MPRAGE Wave-Shuffling and 3D-FSE
Wave-Shuffling data from a single 24 year old healthy male were acquired on a 3T Siemens Prisma
scanner with IRB approval and informed consent.
All data were acquired using the following imaging parameters for brain acquisition: matrix size of
$256 \times 256 \times 256$, 1 mm-isotropic resolution, and 32-channel head coil.
The turbo factor for both MPRAGE and 3D-FSE was set to $256$.
To improve the computational efficiency of the reconstruction, 16-channel SVD coil compression
was applied to all data sets.
Coil sensitivity maps were estimated using ESPIRiT \cite{uecker2014espirit, iyer2020sure} from a
2-second low-resolution GRE calibration scan.
All Shuffling and Wave-Shuffling data were prospectively under-sampled using a variable density
Poisson disc sampling mask generated using the Berkeley Advanced Reconstruction Toolbox (BART)
\cite{bart}.
Each phase encode was sampled at most once.
The Standard Wave MPRAGE data were acquired using a regular $2\times$ under-sampling pattern along
the phase encode $(k_y)$ direction.

\tab
This works lists the total acquisition times $(T_{acq})$ instead of acceleration factors.
An acquisition time of 648 seconds corresponds to a fully sampled non time-resolved
MPRAGE acquisition with the same sequence parameters.

\tab
For all Standard Wave acquisitions, the associated Wave-PSFs were calibrated using the approach
by Cauley et al.\ \cite{cauley2017autocalibrated}.
For the Wave-Shuffling acquisitions, the respective Wave-PSFs were estimated using ShflPSF.
A fifty iteration reconstruction using a single temporal basis (the singular vector associated
with the largest singular value) was used as the initial reconstruction $(\alpha_0)$ in Equation
\eqref{eq:waveopt}.
Nelder-Mead Simplex was used to solve Equation \eqref{eq:waveopt} using the edge-detector $(T)$
defined in Equation \eqref{eq:edgedetector}.
The optimization used ``fminsearch'' in MATLAB (MathWorks, Natick, MA).

\tab
In line with the \nameref{sec:wavetheory} sub-section, the Wave-Shuffling reconstructions were
performed at a synthetic resolution of $\approx 0.71$ mm in the phase encode $(y)$ and partition
encode $(z)$ directions by zero-padding the $k_y$ and $k_z$ dimensions.
The sparsifying transform (the $\Psi$ in Equations \eqref{eq:shflrecon} and \eqref{eq:wshflrecon})
used was locally low-rank (LLR).
The reconstruction algorithm used was the Fast Iterative Soft Thresholding Algorithm (FISTA)
\cite{beck2009fast}.
For all cases, the acquired k-space was normalized to have an $l_2$ norm of one, and
regularization values ($\lambda$) from the below set were searched through for the best
qualitative reconstruction, yielding a regularization value of $\lambda = 0.002$.
$$\lambda \in \{0.0001, 0.0002, \dots, 0.0009, 0.001, 0.002,\dots, 0.009, 0.01, 0.02\}$$
This value was then kept constant over all cases.

\tab
The Wave-Shuffling reconstruction was implemented with BART.
Additionally, BART was compiled using the Intel(R) Math Kernel Library.
The reconstructions were performed on an Intel(R) Xeon(R) Gold 6248R CPU.
\subsubsection{\small \it MPRAGE Experiments} 
As per prior work \cite{polak2019highly}, acquisitions were performed using an inversion
time (TI) of 1100 ms, a repetition time (TR) of 2500 ms and a bandwidth (BW) of 200 Hz/Pixel.
The wave-encoding used a sine wave on one gradient axis and a cosine wave on the other gradient
axis as depicted in Figure \ref{fig:explainer}(A).
The cosine wave-gradient required a ramp up gradient period to reach the highest gradient
amplitude at the start of the acquisition window and a ramp down period after the end of
the acquisition to return to zero.
For simplicity, the current cosine implementation utilized a quarter-cycle shifted sine wave
with an additional half cycle.
This is shown in the $G_z$ gradient of Figure \ref{fig:explainer}(A).
Consequently, when using low wave cycles, the minimum echo spacing increased as the size
of the wave sinusoid was large.
For MPRAGE, at low cycles, this necessitates an increase to the minimum inversion time (TI).
The minimum possible number of cycles achievable while maintaining a TI of 1100 milliseconds
seconds was five.
Consequently, this work used 5 cycles as a lower limit to the number of cycles.

\tab
Standard Wave MPRAGE, MPRAGE Shuffling and MPRAGE Wave-Shuffling data were acquired using the
parameters listed in Table \ref{tbl:params}.
\begin{table}\begin{center} \begin{tabular}{ c | c | c | c | c | c | c} \hline \hline 
Acquisition Type & TI (ms) & TR (ms) & ESP (ms) & Cycles & $G_{\max}$ (mT/m) & $T_{acq}$ (s) \\ \hline \hline 
Standard Wave    & 1100    & 2500    & 7.9      & 17     & 8                 & 324   \\
Standard Wave    & 1100    & 2500    & 8.4      & 5      & 27                & 324   \\
Shuffling        & 1100    & 2500    & 7.8      & 0      & 0                 & 648   \\
Shuffling        & 1100    & 2500    & 7.8      & 0      & 0                 & 230   \\
Shuffling        & 1100    & 2500    & 7.8      & 0      & 0                 & 144   \\
Shuffling        & 1100    & 2500    & 7.8      & 0      & 0                 & 81    \\
Wave-Shuffling   & 1100    & 2500    & 7.9      & 17     & 8                 & 144   \\
Wave-Shuffling   & 1100    & 2500    & 7.9      & 17     & 8                 & 81    \\
Wave-Shuffling   & 1100    & 2500    & 8.4      & 5      & 27                & 144   \\
Wave-Shuffling   & 1100    & 2500    & 8.4      & 5      & 27                & 81    \\ \hline
\end{tabular} \caption{
  MPRAGE acquisition parameters. TI: Inversion Time, TR: Repetition Time.
  ESP: Echo Spacing, $G_{\max}$: Maximum amplitude of the sinusoidal gradients, $T_{acq}$: Acquisition time.
  \label{tbl:params}}\end{center}\end{table}
Regular Shuffling data at various prospective accelerations were acquired to determine how much
acceleration can be achieved.
Wave-Shuffling data at different prospective accelerations were also acquired at 17 cycles with a
$G_{\max}$ of 8 mT/m to match the wave parameters used in prior work for Standard Wave MPRAGE
\cite{polak2019highly}.
Finally, to push $G_{\max}$ of the wave-encoding to the maximum value while keeping the same TI,
Wave-Shuffling data at various prospective accelerations were acquired at 5 cycles with a $G_{\max}$
of 27 mT/m.
To verify the efficacy of the Wave-Shuffling model in mitigating Signal Mixing artifacts (see the
\nameref{sec:waveparamopt} sub-section), Standard Wave MPRAGE data were acquired at 17 cycles with a
$G_{\max}$ of 8 mT/m (to match prior work \cite{polak2019highly}) and at 5 cycles with $G_{\max}$
of 27 mT/m (to demonstrate the Signal Mixing artifacts).

\tab
For each acquisition listed in Table \ref{tbl:params}, the Bloch equation was used to simulate
signal evolutions across the echo train during the inversion recovery of MPRAGE using the
acquisition parameters described in the \nameref{sec:tpsfmethod} sub-section with the
appropriate echo-spacing.

\tab
All reconstructions were allowed to run for 500 FISTA\cite{beck2009fast} iterations.
\subsubsection{\small \it 3D-FSE Experiment.} 
For 3D-FSE, two cycle Wave-Shuffling data were acquired at the maximum possible amplitude of
22 mT/m.
This corresponds to the lowest number of cycles such that the echo spacing and echo train length of
the acquisition is within $20\%$ of prior work \cite{polak2019highly}.
Note that this is a variable flip angle acquisition with stimulated echoes, allowing for the
reconstruction of images at later echo-times compared to conventional spin-echo imaging.
The echo spacing was 4.32 milliseconds, the TR was 3.2 seconds, the BW was 592 Hz per pixel and the
middle echo was acquired at an echo time of 557 milliseconds.
The data was acquired with an acquisition time of 151 seconds.

\tab
The reconstruction was allowed to run for 500 FISTA\cite{beck2009fast} iterations with the
locally low-rank constraint.

\section{Results}
\subsection{Subspace Design}~\\ 
For both MPRAGE and 3D-FSE, a rank of $4$ was determined to be the smallest subspace such that the
NRMSE of any signal in the ensemble was less than $2.5\%$.
This is illustrated in Figure \ref{fig:subspace}, where the x-axis denotes an index into the
signal ensemble dictionary and the y-axis denotes the projection NRMSE.
For MPRAGE, the range (A-B) corresponds to $T_1$ values in range [50, 5000] milliseconds with
a flip angle of $6.3^\circ$,
(B-C) corresponds to the same $T_1$ values with a flip angle of $7.2^\circ$,
(C-D) corresponds to the same $T_1$ values with a flip angle of $8.1^\circ$,
(D-E) corresponds to the same $T_1$ values with a flip angle of $9^\circ$,
(E-F) corresponds to the same $T_1$ values with a flip angle of $9.9^\circ$,
(F-G) corresponds to the same $T_1$ values with a flip angle of $10.8^\circ$, and
(G-H) corresponds to the same $T_1$ values with a flip angle of $11.7^\circ$.
Similarly for 3D-FSE, the range (A-B) corresponds to $T_2$ values in range [20, 2000] milliseconds with
a step size of 2 milliseconds and a $T_1$ of 200 ms,
(B-C) corresponds to the same $T_2$ values with a $T_1$ value of 400 ms,
(C-D) corresponds to the same $T_2$ values with a $T_1$ value of 800 ms,
(D-E) corresponds to the same $T_2$ values with a $T_1$ value of 1400 ms,
(E-F) corresponds to the same $T_2$ values with a $T_1$ value of 2000 ms, and
(F-G) corresponds to the same $T_2$ values with a $T_1$ value of 4000 ms.
\subsection{PSF and Wave Parameter Optimization}~\\
The PSF results are depicted in Figure \ref{fig:tpsf_analysis}.
The maximum of the peak side-lobes as a function of the maximum sinusoid gradient amplitude
$(G_{\max})$ and number of cycles is depicted.
For a particular $G_{\max}$, a lower peak side-lobe is observed when using lower cycles.
Similarly, for a particular cycle, a lower peak side-lobe is observed when using higher $G_{\max}$
values.
Therefore, in the following experiments, the use of Wave-Shuffling acquisitions with a low number of
cycles and large $G_{\max}$ is targeted.

\tab
The PSFs associated with MPRAGE Shuffling (with no wave encoding) and MPRAGE Wave-Shuffling with
the highest $G_{\max}$ of 27 mT/m at 5 cycles are plotted.
The maximum intensity projection along the $(x, y), (x, z)$ and $(y, z)$ spatial axes are taken
to better visualize the incoherence.
``\customsubscript{PSF}{1}'' and ``\customsubscript{PSF}{2}'' depict the Shuffling outputs of $\delta_1$
and $\delta_2$ respectively.
Similarly, ``\customsubscript{WPSF}{1}'' and ``\customsubscript{WPSF}{2}'' depict the Wave-Shuffling outputs
of $\delta_1$ and $\delta_2$ respectively.
The green dashed lines denote the Shuffling results and the red solid lines denote the
Wave-Shuffling results.
The voxel spreading induced by the sinusoidal wave-gradients in Wave-Shuffling are seen to
distribute the incoherent artifacts across all the volumetric dimensions $(x, y, z)$ resulting
in lower peak side-lobes.
In particular, the second coefficient of \customsubscript{PSF}{1}/\customsubscript{WPSF}{1} and the first
coefficient of \customsubscript{PSF}{2}/\customsubscript{WPSF}{2} demonstrates how Wave-Shuffling achieves
less signal cross-talk between coefficients.
\subsection{In-Vivo Experiments}
\subsubsection{\small \it MPRAGE Experiments.} 
Figure \ref{fig:tfl_shfl_vs_wshfl} depicts the resulting reconstructions at selected TI times for
MPRAGE Shuffling at various levels of acceleration with $T_{acq}$ in seconds of $(648, 230, 144)$, and
MPRAGE Wave-Shuffling at high acceleration with $T_{acq} = 81 s$.
While the early TIs have lower signal levels than the later TIs, for ease of visualization,
each TI image is individually normalized.
This figure illustrates how Wave-Shuffling provides comparable reconstruction
at $\approx8\times$ higher acceleration compared to Shuffling.
In particular, MPRAGE Shuffling is stable at $T_{acq} = 648s$ but suffers from reconstruction
artifacts at higher accelerations that are particularly noticeable at the early time points.
In contrast, MPRAGE Wave-Shuffling at $T_{acq} = 81s$ is able to stably recover the underlying time series
with reconstruction quality comparable to MPRAGE Shuffling at $T_{acq} = 648s$ up-to noise considerations
arising from high levels of acceleration.

\tab
Figure \ref{fig:wshfl_vs_accel} demonstrates the reconstruction quality of Wave-Shuffling as a
function of wave-encoding parameters, where reconstructions from a low-SNR
time-point with low signal level ($TI$ of 430 ms) are shown to highlight the differences
in reconstruction performance.
Regular Shuffling (no wave-encoding) is seen to have severe artifacts at the displayed time
point at both accelerations $T_{acq} = 144 s$ and $T_{acq} = 81s$.
The 17-cycle, $G_{\max} = 8$ mT/m case is seen to improve reconstruction at $T_{acq} = 144 s$ and
$T_{acq} = 81s$.
The 5-cycle, $G_{\max} = 27$ mT/m case is seen to provide much better conditioning and stably
recovers the data at $T_{acq} = 81s$.

\tab
Figures \ref{fig:tfl_shfl_vs_wshfl} and \ref{fig:wshfl_vs_accel} demonstrate how through the use of
optimized wave-encoding, Wave-Shuffling was able to achieve significant increase in acceleration.
Note that there are slight contrast differences between the reconstructed image at the specific
inversion times due to echo spacing differences between the acquisitions (listed in Table
\ref{tbl:params}).

\tab
Figure \ref{fig:signal_mixing} shows how the Shuffling model is able to mitigate Signal Mixing
artifacts.
Standard Wave MPRAGE with parameters used as per prior work \cite{polak2019highly} at 17 cycles,
$G_{\max} = 8$ mT/m shows good reconstruction with no artifacts.
However, Standard Wave MPRAGE at 5 cycles, $G_{\max} = 27$ mT/m suffers from significant
ringing artifacts arising from the signal recovery mixing across the echo-train during
partition-encoding.
In contrast, Wave-Shuffling at the same wave parameters (5 cycles, $G_{\max}=27$ mT/m) at the
comparable TI of 1100 milliseconds shows no ringing artifacts even at high acceleration
$(T_{acq} = 81s)$.
This confirms that modelling the temporal evolution over the echo-train avoids the Signal Mixing.

\tab
Figure \ref{fig:psf_calibration} demonstrates how ShflPSF is able to correct for gradient hardware
errors and mitigate ringing artifacts related to incorrect wave calibration.
This particular figure depicts ShflPSF applied to the MPRAGE Wave-Shuffling at $T_{acq} = 81s$ with
5 cycles and a $G_{\max}$ of 27 mT/m.
ShflPSF is able to correct for gradient hardware errors and mitigate ringing artifacts related to
incorrect wave sinusoids by performing a post-processing deconvolution with $W_e^*$ as described in
the \nameref{sec:autopsf} subsection.

\tab
Figure \ref{fig:tfl_showcase} represents a showcase of the Wave-Shuffling method when applied to
the MPRAGE sequence for 1-mm isotropic resolution at $T_{acq} = 81s$.
Four out of the 256 reconstructed images are depicted.
A highly accelerated MPRAGE Wave-Shuffling acquisition and reconstruction is seen to stably recover
a time series of images with multiple contrasts at high quality up to noise considerations.
Note that the CSF is hypo-dense compared to relative tissue, particularly at short TE.
Due to having a high $T_1$ value, the CSF is unable to fully recover within the prescribed TR
and consequently, once the signal reaches steady-state, the CSF signal is suppressed relative
to the surrounding tissue.
\subsubsection{\small \it 3D-FSE Experiment.} 
Figure \ref{fig:tse_showcase} represents a showcase of the Wave-Shuffling method when applied to
the 3D-FSE sequence for 1-mm isotropic acquisition at $T_{acq} = 151 s$.
Four out of the 256 reconstructed images are depicted.
A highly accelerated 3D-FSE Wave-Shuffling acquisition and reconstruction is seen to stably
recover a time series of images with multiple contrasts.
Since 3D-FSE uses a variable flip angle train to stimulate echoes, there is still signal at
around 500 ms.

\section{Discussion}
In this work, the Wave-Shuffling technique was developed for fast time-resolved structural
imaging, where wave-encoding parameters were optimized using the PSF analysis to improve
acceleration capability.
Wave-Shuffling was successfully implemented on MPRAGE and 3D-FSE sequences and demonstrated
to provide fast high-quality time-resolved brain imaging.

\tab
A direct application of the Wave-Shuffling technique is the further reduction of scan
times for knee MRI exams \cite{tamir2019}.
A promising direction of Wave-Shuffling MPRAGE is to leverage the multiple-contrasts
to study deep-brain structure like the lateral geniculate nucleus (LGN) which is more
visible at the earlier TIs \cite{datta2021}.
This is motivated by the quality of the reconstruction performance demonstrated by
Figure \ref{fig:tfl_showcase} with just $81$ seconds of acquisition.
In particular, the SNR is expected to significantly improve at scan times comparable
to MP2RAGE \cite{marques2010, mussard2020}.
Note that Wave-Shuffling MPRAGE as discussed in this work resolves 256 echo-images
over the 2 echo-images recovered by MP2RAGE.

\tab
With respect to Figure \ref{fig:subspace}, for MPRAGE, note that the SVD-derived subspace has
difficulty representing low $T_1$ values at lower ranks.
For 3D-FSE, a similar pattern is seen for low $T_2$ values.
While additional basis vectors results in lower projection NRMSE, the added unknowns
ill-conditions the reconstructions described by Equations \eqref{eq:shflrecon} and
\eqref{eq:wshflrecon}.

\tab
For MPRAGE Wave-Shuffling, the lower bound on the number of wave cycles was set empirically
to five to maintain a TI of 1100 ms.
A four cycle acquisition corresponds to a TI of 1130 ms and a three cycle acquisition
corresponds to a TI of 1170 ms.
Lower wave cycle numbers can achieve higher $G_{\max}$ but may result in differential phase
modulation within a voxel resulting in intra-voxel de-phasing \cite{bilgic2015wave},
and at three cycles, the MPRAGE Wave-Shuffling reconstruction suffers from significant
artifacts.
With these constraints in mind, the PSF results in Figure \ref{fig:tpsf_analysis} are seen to
be a good qualitative indicator of Wave-Shuffling reconstruction performance.
Similarly, for 3D-FSE, one cycle wave-encoding results in the non-trivial artifacts.

\tab
The reconstruction noise level is tied to the amount of acceleration.
Wave-Shuffling, being a time-resolving method, is encoding-limited and SNR-limited (due to
less noise averaging) at high accelerations, and consequently the reconstructions presented in
Figures \ref{fig:tfl_shfl_vs_wshfl}, \ref{fig:wshfl_vs_accel} and \ref{fig:signal_mixing} show
increased noise levels.
Areas to explore in this regard are optimized flip-angle encoding for a more noise-performant
acquisition and machine-learning based de-noising.

\tab
ShflPSF is seen to mitigate the ringing artifacts associated with an incorrect Wave-PSF.
Additionally, the method is flexible enough to be incorporated into model-based Wave-encoding
reconstruction.
It is empirically observed that a single coefficient usually suffices to effectively correct
for the ringing artifacts.
Since the Fourier coefficient is allowed to be complex, the magnitude of the Fourier
coefficient effectively models gradient amplitude errors while the phase of the
coefficient encodes gradient delay errors.

\tab
Multiple rounds of ShflPSF can be performed, where the optimization alternates between
Wave-PSF calibration and Wave-Shuffling reconstruction.
However, this is not seen to noticeably improve calibration or reconstruction performance
while increasing the computational burden of the method.
Additionally, for Standard Wave acquisitions, ShflPSF performs similarly to the method
by Cauley et al.\ \cite{cauley2017autocalibrated}.

\tab
MPRAGE and 3D-FSE are commonly used sequences for $T_1-$ and $T_2-$ weighted images
respectively.
These sequences are designed for maximizing contrasts differences between tissues of
interest and are not focused on quantitative mapping.
In particular, the signal evolutions over the echo train are highly correlated for
both sequences.
To give a specific example, for the MPRAGE sequence, assuming a flip angle of $9^\circ$,
the absolute correlation coefficient between two simulated signals with $T_1$ values of
800 ms and 1000 ms is $\approx0.99$.
Similarly, for 3D-FSE, assuming a $T_1$ of 800 ms, the absolute correlation coefficient
between two simulated signals with $T_2$ values of 20 ms and 40 ms is $\approx0.98$.
Since this work was focused on augmenting existing sequences with known contrasts
without modifying sequence parameters, it is not optimized for $T_1$ and $T_2$
mapping as there is no clean separability between the highly correlated signal
evolutions.
Future work would be to modify the flip angle train and other sequence parameters to
increase the separability of the signal evolutions of tissues of interest in order to
better fit for quantitative maps.
That being said, to demonstrate what can be achieved without any sequence modifications,
$T_1$ dictionary fitting for Wave-Shuffling MPRAGE is presented in the Supplementary
material.

\tab
Given the flexibility of the wave-encoding and Shuffling models, Wave-Shuffling can be easily
extended to numerous other sequences with the possibility of complimentary-sampling and
joint-reconstruction across multiple Wave-Shuffling acquisitions for higher rates of acceleration
and better noise performance \cite{bilgic2018improving}, which may be further enhanced with Deep
Learning methods \cite{polak2020joint, hammernik2018learning, aggarwal2019modl}.
For further improvement in reconstruction performance, sampling mask optimization in a manner
that is synergistic with the chosen regularization can be utilized \cite{haldar2019oedipus}.

\section{Conclusion}

Wave-encoding better conditions Shuffling reconstruction, enabling higher rates of acceleration.
ShflPSF enables auto-calibrating Wave-PSF calibration for Wave-Shuffling acquisitions.
Wave-Shuffling is applied to MPRAGE and 3D-FSE to achieve rapid, 1 mm-isotropic resolution,
multi-contrast, time-resolved imaging of the brain.

\section{Acknowledgements}

We would like to thank the anonymous referees for their help in improving this manuscript, and
Yamin Arefeen for reading through this manuscript.

\section{Data Availability Statement}

The data and code used to generate Figure \ref{fig:tfl_showcase} are openly available at
\url{https://github.com/sidward/wave-shuffling} with DOI:
\url{https://doi.org/10.5281/zenodo.4603207}.
This repository depends on C code available as a part of the BART Toolbox at
\url{https://github.com/mrirecon/bart/} with DOI:
\url{https://doi.org/10.5281/zenodo.592960}.

\newpage
\bibliographystyle{mrm}
\bibliography{main}

\begin{thebibliography}{10}

\bibitem{mugler2014optimized}
{{Mugler III}}~JP.
\newblock {Optimized three-dimensional fast-spin-echo MRI}.
\newblock Journal of Magnetic Resonance Imaging 2014; 39:745--767.

\bibitem{muglermp}
{{Mugler III}}~JP, Brookeman~JR.
\newblock {Three-dimensional magnetization-prepared rapid gradient-echo imaging
  (3D MP RAGE)}.
\newblock {Magnetic Resonance in Medicine} 1990; 15:152--157.

\bibitem{4193454}
{Liang}~Z.
\newblock {Spatiotemporal Imaging with Partially Separable Functions}.
\newblock { In:} 4th IEEE International Symposium on Biomedical Imaging: From
  Nano to Macro, Virginia, USA, 2007.  pp. 988--991.

\bibitem{zhao2015accelerated}
Zhao~B, Lu~W, Hitchens~TK, Lam~F, Ho~C, Liang~ZP.
\newblock {Accelerated MR Parameter Mapping with Low-rank and Sparsity
  Constraints}.
\newblock {Magnetic Resonance in Medicine} 2015; 74:489--498.

\bibitem{tamir2017t2}
Tamir~JI, Uecker~M, Chen~W, Lai~P, Alley~MT, Vasanawala~SS, Lustig~M.
\newblock {$T_2-$Shuffling: Sharp, Multicontrast, Volumetric Fast Spin-echo
  Imaging}.
\newblock Magnetic Resonance in Medicine 2017; 77:180--195.

\bibitem{tamir2019}
Tamir~JI, Taviani~V, Alley~MT, Perkins~BC, Hart~L, O'Brien~K, Wishah~F,
  Sandberg~JK, Anderson~MJ, Turek~JS, Willke~TL, Lustig~M, Vasanawala~SS.
\newblock {Targeted rapid knee MRI exam using $T_2$-Shuffling}.
\newblock {Journal of Magnetic Resonance Imaging} 2019; 49:e195--e204.

\bibitem{bao2017fast}
Bao~S, Tamir~JI, Young~JL, Tariq~U, Uecker~M, Lai~P, Chen~W, Lustig~M,
  Vasanawala~SS.
\newblock {Fast Comprehensive Single-sequence Four-dimensional Pediatric Knee
  MRI with T2 Shuffling}.
\newblock Journal of Magnetic Resonance Imaging 2017; 45:1700--1711.

\bibitem{cao2018shuffled}
Cao~P, Zhu~X, Tang~S, Leynes~A, Jakary~A, Larson~PE.
\newblock {Shuffled Magnetization-prepared Multicontrast Rapid Gradient-echo
  Imaging}.
\newblock Magnetic Resonance in Medicine 2018; 79:62--70.

\bibitem{kecskemeti2016mpnrage}
Kecskemeti~S, Samsonov~A, Hurley~SA, Dean~DC, Field~A, Alexander~AL.
\newblock {MPnRAGE: A Technique to Simultaneously Acquire Hundreds of
  Differently Contrasted MPRAGE Images with Applications to Quantitative T1
  Mapping}.
\newblock Magnetic Resonance in Medicine 2016; 75:1040--1053.

\bibitem{wang2019echo}
Wang~F, Dong~Z, Reese~TG, Bilgic~B, Manhard~MK, Chen~J, Polimeni~JR, Wald~LL,
  Setsompop~K.
\newblock Echo planar time-resolved imaging {{(EPTI)}}.
\newblock {Magnetic Resonance in Medicine} 2019; 81:3599--3615.

\bibitem{dong2020echo}
Dong~Z, Wang~F, Reese~TG, Bilgic~B, Setsompop~K.
\newblock Echo planar time-resolved imaging with subspace reconstruction and
  optimized spatiotemporal encoding.
\newblock {Magnetic Resonance in Medicine} 2020; 84:2442--2455.

\bibitem{bilgic2015wave}
Bilgic~B, Gagoski~BA, Cauley~SF, Fan~AP, Polimeni~JR, Grant~PE, Wald~LL,
  Setsompop~K.
\newblock {Wave-CAIPI for Highly Accelerated 3D Imaging}.
\newblock Magnetic Resonance in Medicine 2015; 73:2152--2162.

\bibitem{polak2019highly}
Polak~D, Cauley~S, Huang~SY, Longo~MG, Conklin~J, Bilgic~B, Ohringer~N,
  Raithel~E, Bachert~P, Wald~LL, Setsompop~K.
\newblock {Highly-accelerated Volumetric Brain Examination Using Optimized
  Wave-CAIPI Encoding}.
\newblock Journal of Magnetic Resonance Imaging 2019; 50:961--974.

\bibitem{polak2018wave}
Polak~D, Setsompop~K, Cauley~SF, Gagoski~BA, Bhat~H, Maier~F, Bachert~P,
  Wald~LL, Bilgic~B.
\newblock {Wave-CAIPI for Highly Accelerated MP-RAGE Imaging}.
\newblock Magnetic Resonance in Medicine 2018; 79:401--406.

\bibitem{moriguchi2006bunched}
Moriguchi~H, Duerk~JL.
\newblock {Bunched Phase Encoding (BPE): A New Fast Data Acquisition Method in
  MRI}.
\newblock Magnetic Resonance in Medicine 2006; 55:633--648.

\bibitem{breuer2005controlled}
Breuer~FA, Blaimer~M, Heidemann~RM, Mueller~MF, Griswold~MA, Jakob~PM.
\newblock {Controlled Aliasing in Parallel Imaging Results in Higher
  Acceleration (CAIPIRINHA) for Multi-slice Imaging}.
\newblock Magnetic Resonance in Medicine 2005; 53:684--691.

\bibitem{breuer2006controlled}
Breuer~FA, Blaimer~M, Mueller~MF, Seiberlich~N, Heidemann~RM, Griswold~MA,
  Jakob~PM.
\newblock {Controlled Aliasing in Volumetric Parallel Imaging (2D CAIPIRINHA)}.
\newblock Magnetic Resonance in Medicine 2006; 55:549--556.

\bibitem{bilgic2016optimized}
Bilgic~B, Ye~H, Wald~L, Setsompop~K.
\newblock {Optimized CS-wave Imaging with Tailored Sampling and Efficient
  Reconstruction}.
\newblock Proceedings of the 24th Annual Meeting ISMRM 2016; p. 612.

\bibitem{chen2018self}
Chen~F, Taviani~V, Tamir~JI, Cheng~JY, Zhang~T, Song~Q, Hargreaves~BA,
  Pauly~JM, Vasanawala~SS.
\newblock Self-calibrating wave-encoded variable-density single-shot fast spin
  echo imaging.
\newblock {Journal of Magnetic Resonance Imaging} 2018; 47:954--966.

\bibitem{cauley2017autocalibrated}
Cauley~SF, Setsompop~K, Bilgic~B, Bhat~H, Gagoski~B, Wald~LL.
\newblock {Autocalibrated Wave-CAIPI Reconstruction: Joint Optimization of
  k-space Trajectory and Parallel Imaging Reconstruction}.
\newblock {Magnetic Resonance in Medicine} 2017; 78:1093--1099.

\bibitem{ye2016accelerating}
Ye~H, Ma~D, Jiang~Y, Cauley~SF, Du~Y, Wald~LL, Griswold~MA, Setsompop~K.
\newblock {Accelerating magnetic resonance fingerprinting (MRF) using t-blipped
  simultaneous multislice (SMS) acquisition}.
\newblock {Magnetic Resonance in Medicine} 2016; 75:2078--2085.

\bibitem{weigel2015extended}
Weigel~M.
\newblock {Extended Phase Graphs: Dephasing, RF Pulses, and Echoes - Pure and
  Simple}.
\newblock Journal of Magnetic Resonance Imaging 2015; 41:266--295.

\bibitem{uecker2014espirit}
Uecker~M, Lai~P, Murphy~MJ, Virtue~P, Elad~M, Pauly~JM, Vasanawala~SS,
  Lustig~M.
\newblock {ESPIRiT - An eigenvalue approach to autocalibrating parallel MRI:
  where SENSE meets GRAPPA}.
\newblock Magnetic Resonance in Medicine 2014; 71:990--1001.

\bibitem{iyer2020sure}
Iyer~S, Ong~F, Setsompop~K, Doneva~M, Lustig~M.
\newblock Sure-based automatic parameter selection for {{ESPIRiT}} calibration.
\newblock {Magnetic Resonance in Medicine} 2020; 84:3423--3437.

\bibitem{bart}
Uecker~M, Ong~F, Tamir~JI, Bahri~D, Virtue~P, Cheng~JY, Zhang~T, Lustig~M.
\newblock {Berkeley Advanced Reconstruction Toolbox}.
\newblock Proc. Intl. Soc. Mag. Reson. Med 2015; 23.

\bibitem{beck2009fast}
Beck~A, Teboulle~M.
\newblock {A Fast Iterative Shrinkage-thresholding Algorithm for Linear Inverse
  Problems}.
\newblock SIAM Journal on Imaging Sciences 2009; 2:183--202.

\bibitem{datta2021}
Datta~R, Bacchus~MK, Kumar~D, Elliott~MA, Rao~A, Dolui~S, Reddy~R, Banwell~BL,
  Saranathan~M.
\newblock {Fast automatic segmentation of thalamic nuclei from MP2RAGE
  acquisition at 7 Tesla}.
\newblock {Magnetic Resonance in Medicine} 2021; 85:2781--2790.

\bibitem{marques2010}
Marques~JP, Kober~T, Krueger~G, van~der Zwaag~W, Van~de Moortele~PF,
  Gruetter~R.
\newblock {MP2RAGE: A self bias-field corrected sequence for improved
  segmentation and $T_1$-mapping at high field}.
\newblock Neuroimage 2010; 49:1271--1281.

\bibitem{mussard2020}
Mussard~E, Hilbert~T, Forman~C, Meuli~R, Thiran~JP, Kober~T.
\newblock {Accelerated MP2RAGE imaging using Cartesian phyllotaxis readout and
  compressed sensing reconstruction}.
\newblock {Magnetic Resonance in Medicine} 2020; 84:1881--1894.

\bibitem{bilgic2018improving}
Bilgic~B, Kim~TH, Liao~C, Manhard~MK, Wald~LL, Haldar~JP, Setsompop~K.
\newblock {Improving parallel imaging by jointly reconstructing multi-contrast
  data}.
\newblock {Magnetic Resonance in Medicine} 2018; 80:619--632.

\bibitem{polak2020joint}
Polak~D, Cauley~S, Bilgic~B, Gong~E, Bachert~P, Adalsteinsson~E, Setsompop~K.
\newblock {Joint multi-contrast variational network reconstruction (jVN) with
  application to rapid 2D and 3D imaging}.
\newblock {Magnetic Resonance in Medicine} 2020; 84:1456--1469.

\bibitem{hammernik2018learning}
Hammernik~K, Klatzer~T, Kobler~E, Recht~MP, Sodickson~DK, Pock~T, Knoll~F.
\newblock {Learning a variational network for reconstruction of accelerated MRI
  data}.
\newblock {Magnetic Resonance in Medicine} 2018; 79:3055--3071.

\bibitem{aggarwal2019modl}
{Aggarwal}~HK, {Mani}~MP, {Jacob}~M.
\newblock {MoDL: Model-Based Deep Learning Architecture for Inverse Problems}.
\newblock {IEEE Transactions on Medical Imaging} 2019; 38:394--405.

\bibitem{haldar2019oedipus}
{Haldar}~JP, {Kim}~D.
\newblock {OEDIPUS: An Experiment Design Framework for Sparsity-Constrained
  MRI}.
\newblock {IEEE Transactions on Medical Imaging} 2019; 38:1545--1558.

\end{thebibliography}
\clearpage

\newpage
\fig{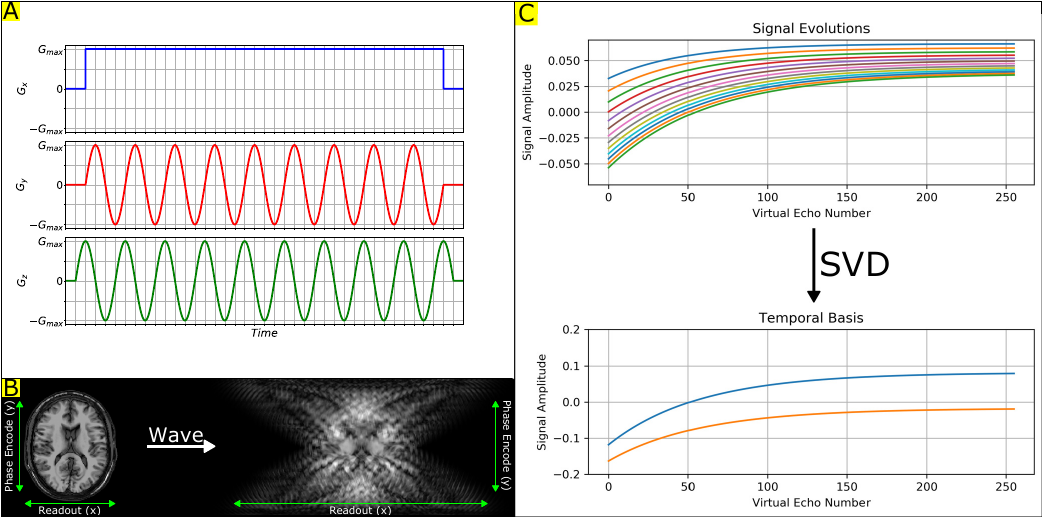}{0.98}{fig:explainer}{
  (A) In wave-encoding, additional sinusoidal $(G_y, G_z)$ gradients are applied during the readout.
  (B) The sinusoidal gradients induces a voxel spreading effect along the readout direction.
      The amount of spreading increases linearly as a function of the spatial $(y, z)$ locations.
  (C) This depicts simulated MPRAGE signal evolutions.
      These realistic signal evolutions are observed to live in a low-rank subspace that can be
      obtained via the Singular Value Decomposition (SVD).
}

\newpage

\fig{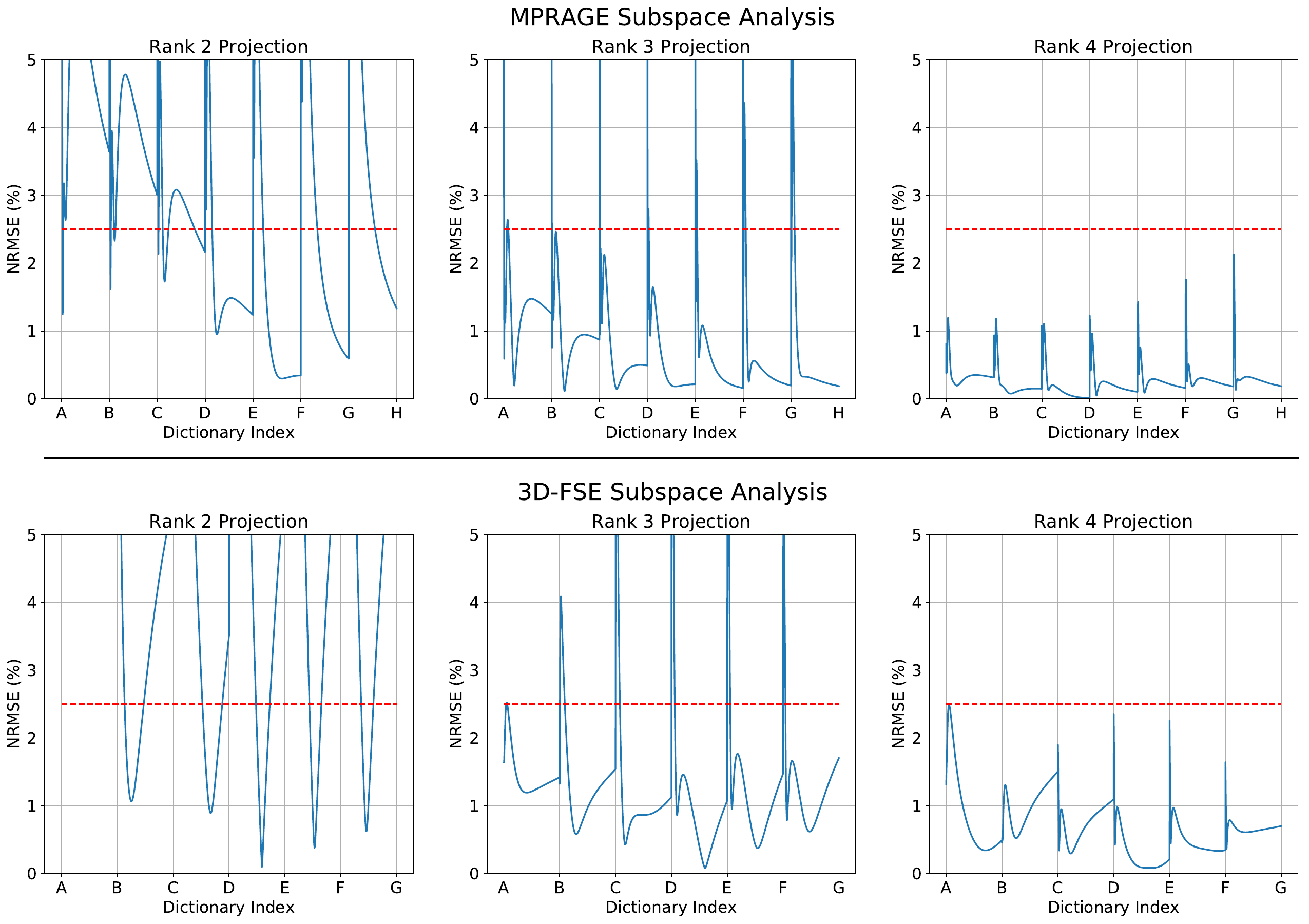}{0.98}{fig:subspace}{
  The simulated signal ensemble for MPRAGE and 3D-FSE are projected onto SVD-derived subspaces
  of varying ranks and the NRMSE per signal is calculated.
  The red dotted line denotes the 2.5\% cut-off.
  For MPRAGE, the range (A-B) corresponds to $T_1$ values in range [50, 5000] milliseconds with
  a flip angle of $6.3^\circ$,
  (B-C) corresponds to the same $T_1$ values with a flip angle of $7.2^\circ$,
  (C-D) corresponds to the same $T_1$ values with a flip angle of $8.1^\circ$,
  (D-E) corresponds to the same $T_1$ values with a flip angle of $9^\circ$,
  (E-F) corresponds to the same $T_1$ values with a flip angle of $9.9^\circ$,
  (F-G) corresponds to the same $T_1$ values with a flip angle of $10.8^\circ$, and
  (G-H) corresponds to the same $T_1$ values with a flip angle of $11.7^\circ$.
  For 3D-FSE, the range (A-B) corresponds to $T_2$ values in range [20, 2000] milliseconds with
  a $T_1$ of 200 ms,
  (B-C) corresponds to the same $T_2$ values with a $T_1$ value of 400 ms,
  (C-D) corresponds to the same $T_2$ values with a $T_1$ value of 800 ms,
  (D-E) corresponds to the same $T_2$ values with a $T_1$ value of 1400 ms,
  (E-F) corresponds to the same $T_2$ values with a $T_1$ value of 2000 ms, and
  (F-G) corresponds to the same $T_2$ values with a $T_1$ value of 4000 ms.
}

\newpage

\fig{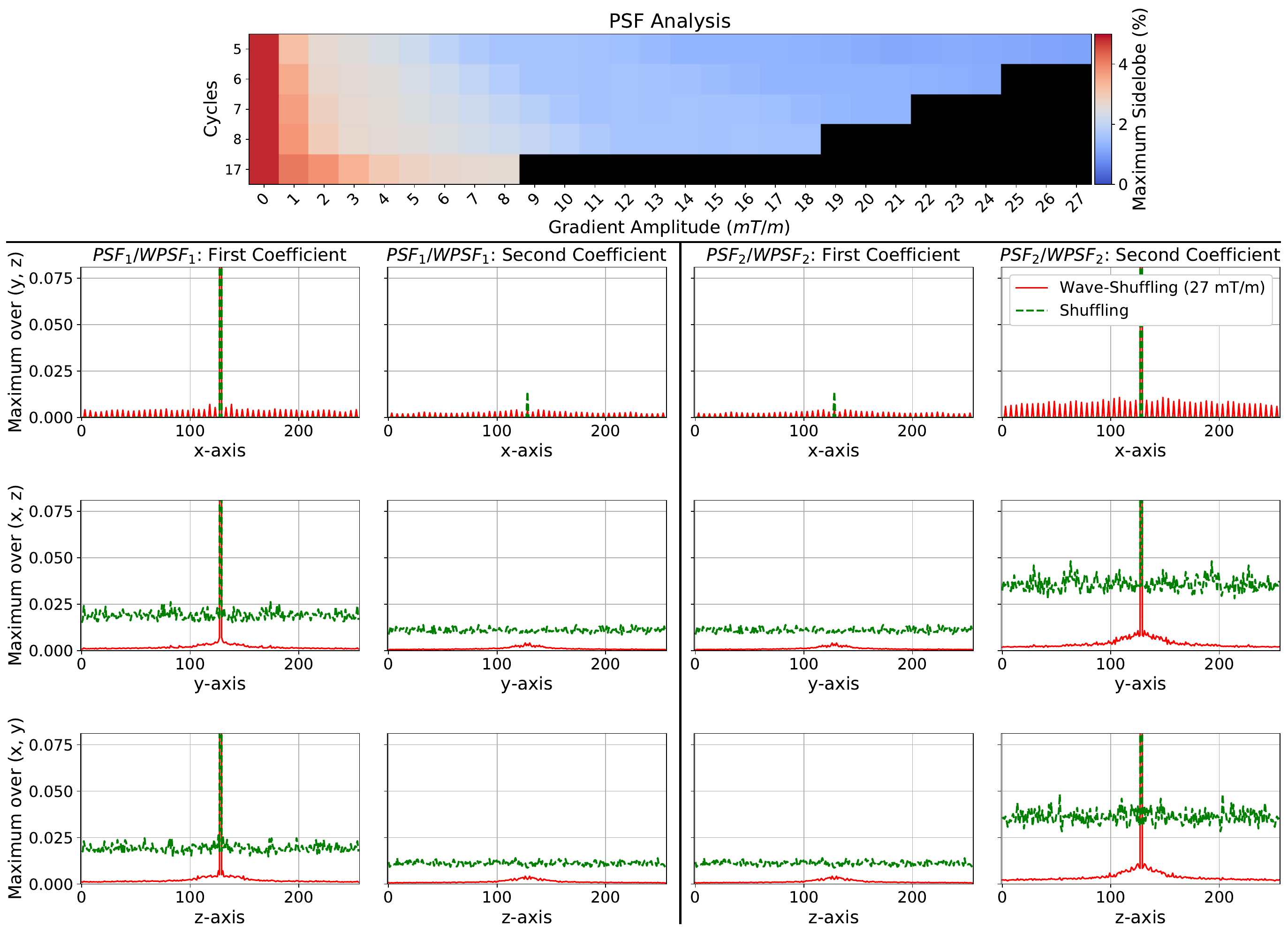}{0.98}{fig:tpsf_analysis}{
  (Top) The maximum peak side-lobe as a function of sinusoid maximum gradient amplitude
  $(G_{\max})$ and number of cycles is depicted.
  (See Equation \eqref{eq:tpsfcost}.)
  A lower peak side-lobe is desirable.
  The black color denotes unreachable parameters due to gradient hardware slew rate limitations.
  (Bottom Left and Bottom Right)
  The PSFs associated with MPRAGE Shuffling (with no wave encoding) and high
  $G_{\max}$ MPRAGE Wave-Shuffling are plotted for comparison.
  The former is plotted with a dotted green line, while the latter is plotted with a solid red
  line.
  The legend of the PSFs plots are depicted in the top-right sub-plot.
  The left two columns denote the first coefficient response while the right two columns denote
  the second coefficient response.
  To better visualize the side lobes, the peaks of the respective PSFs are scaled to one
  and the range $[0, 0.1]$ is plotted.
  The maximum intensity projection along the $(x, y), (x, z)$ and $(y, z)$ spatial axes are taken
  to better visualize the incoherence.
  \customsubscript{PSF}{1}/\customsubscript{WPSF}{1} and \customsubscript{PSF}{2}/\customsubscript{WPSF}{2} depict the
  Shuffling/Wave-Shuffling outputs of $\delta_1$ and $\delta_2$ respectively.
}

\newpage

\fig{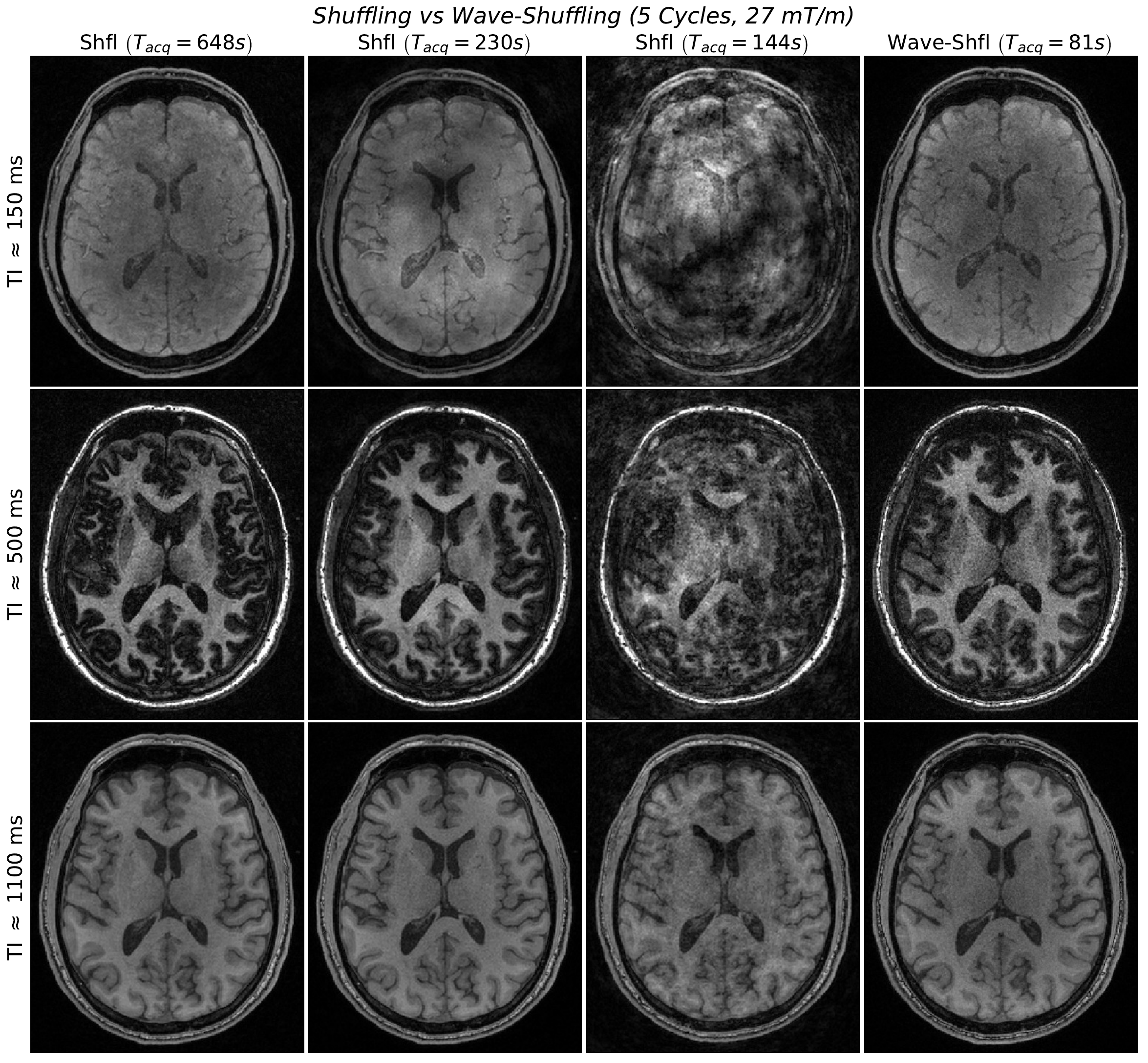}{0.98}{fig:tfl_shfl_vs_wshfl}{
  This figure depicts MPRAGE Shuffling at various accelerations and MPRAGE Wave-Shuffling at high
  acceleration.
  MPRAGE Shuffling is stable at $T_{acq} = 648 s$ but suffers from reconstruction artifacts at lower acquisition times.
  MPRAGE Wave-Shuffling at a significantly lower acquisition time provides well conditioned reconstruction with
  image quality comparable to MPRAGE Shuffling at $T_{acq}=648 s$ up-to noise considerations due to sub-sampling.
}

\newpage

\fig{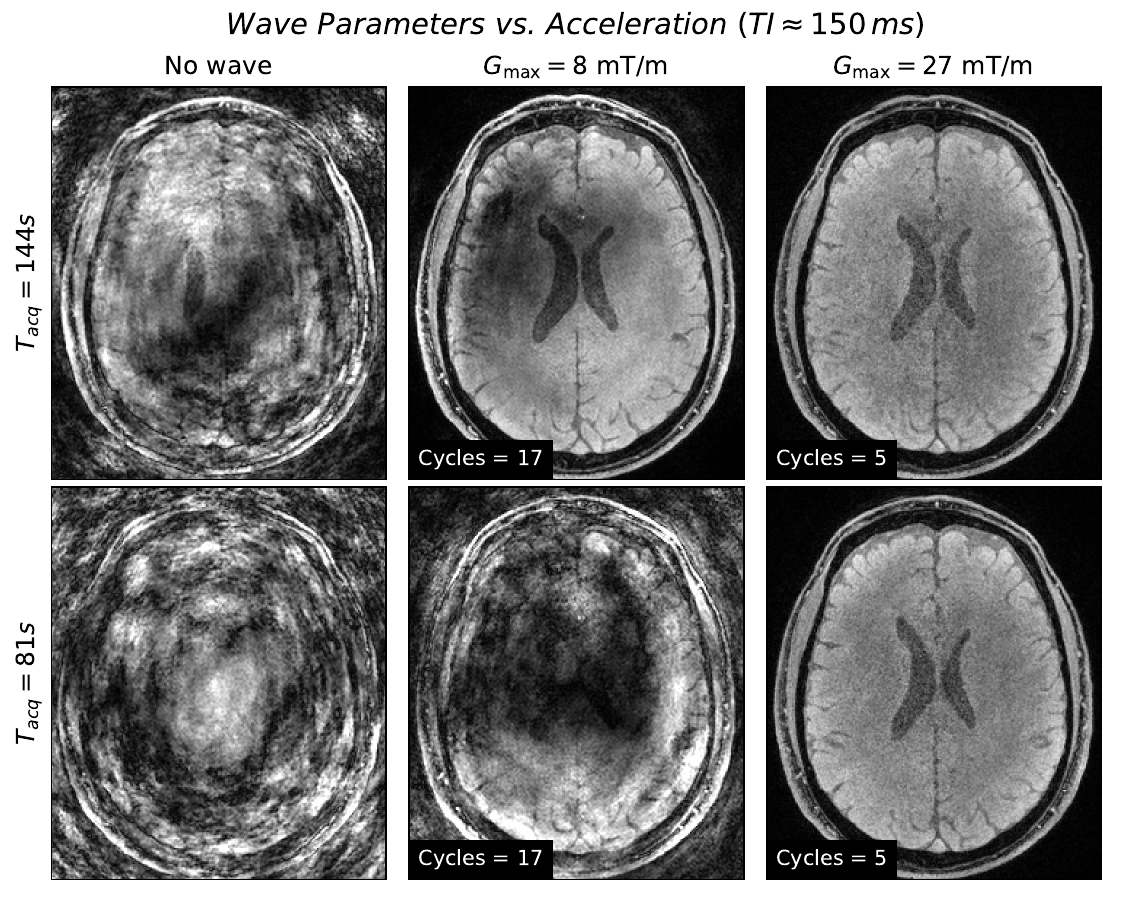}{0.98}{fig:wshfl_vs_accel}{
  The reconstruction quality of Wave-Shuffling varies as a function of wave-encoding parameters.
  Regular Shuffling is seen to have severe artifacts at the hard-to-reconstruct displayed time point
  (TI = 430 ms) at both accelerations ($T_{acq} = 144 s$ and $T_{acq} = 81 s$).
  The 5-cycle, $G_{\max}$ of 27 mT/m case demonstrates the superior encoding capability of high
  $G_{\max}$ wave-encoding with significantly better reconstruction quality at both accelerations.
}

\newpage

\fig{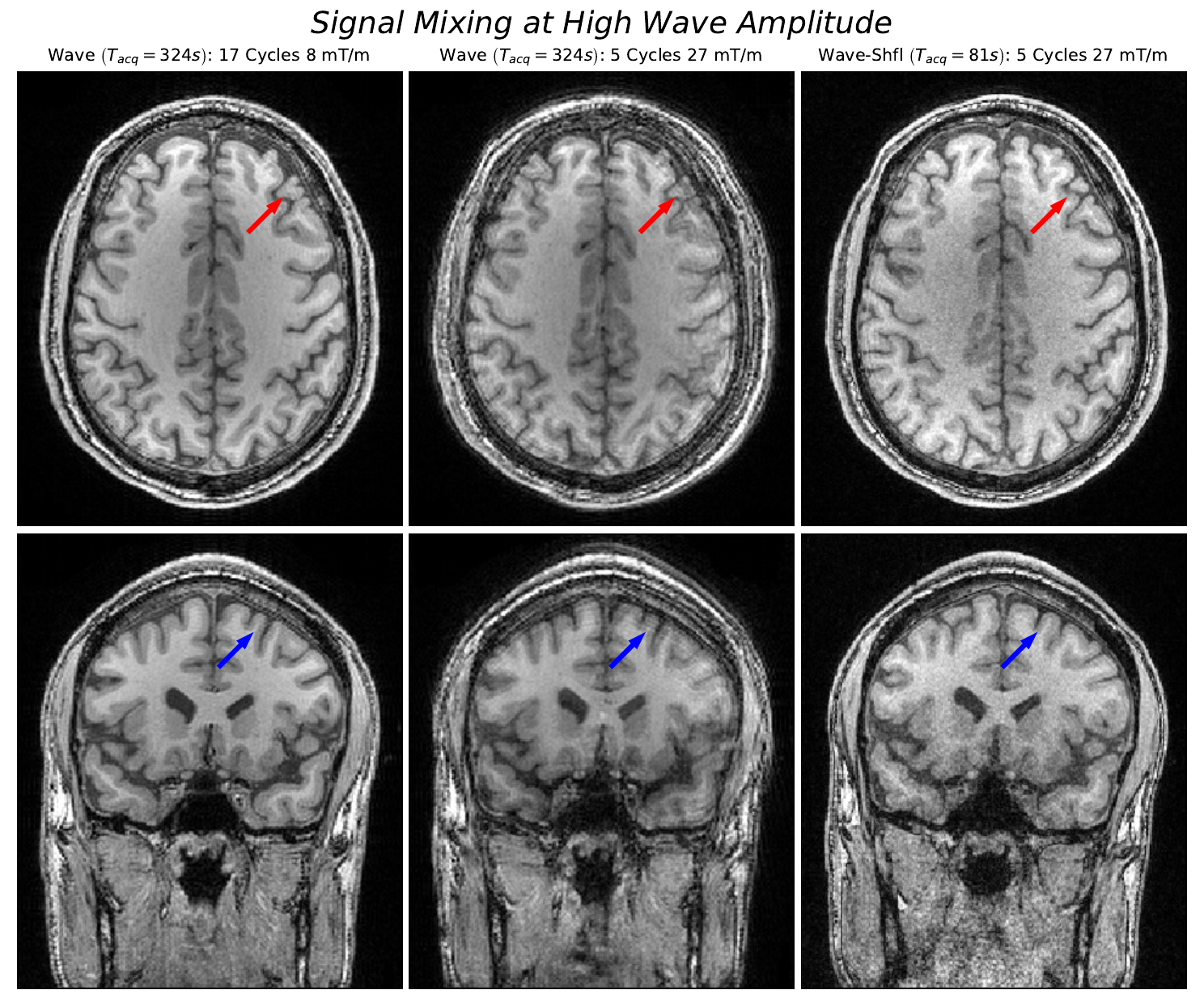}{0.98}{fig:signal_mixing}{
  The inclusion of the Shuffling model is seen to mitigate Signal Mixing artifacts associated
  with high gradient amplitude $(G_{\max})$ wave-encoding acquisitions.
  Standard Wave MPRAGE at 17 cycles, $G_{\max}$ of 8 mT/m shows good reconstruction with
  no artifacts.
  Standard Wave MPRAGE at 5 cycles, $G_{\max}$ of 27 mT/m suffers from significant
  ringing artifacts due to signal recovery over the partition encode direction.
  MPRAGE Wave-Shuffling at the same parameters (5 cycles, $G_{\max}$ of 27 mT/m) at the comparable
  TI of 1100 milliseconds shows no ringing artifacts even at high acceleration.
  For high $G_{\max}$ Standard Wave MPRAGE (the middle column), the red and blue arrows highlight two different
  locations where the Signal-Mixing artifacts are observed.
  For the other two acquisitons, the arrows point to the respective comparable locations to demonstrate that
  low $G_{\max}$ Standard Wave MPRAGE and high $G_{\max}$ MPRAGE Wave-Shuffling are robust to said artifacts.
}

\newpage

\fig{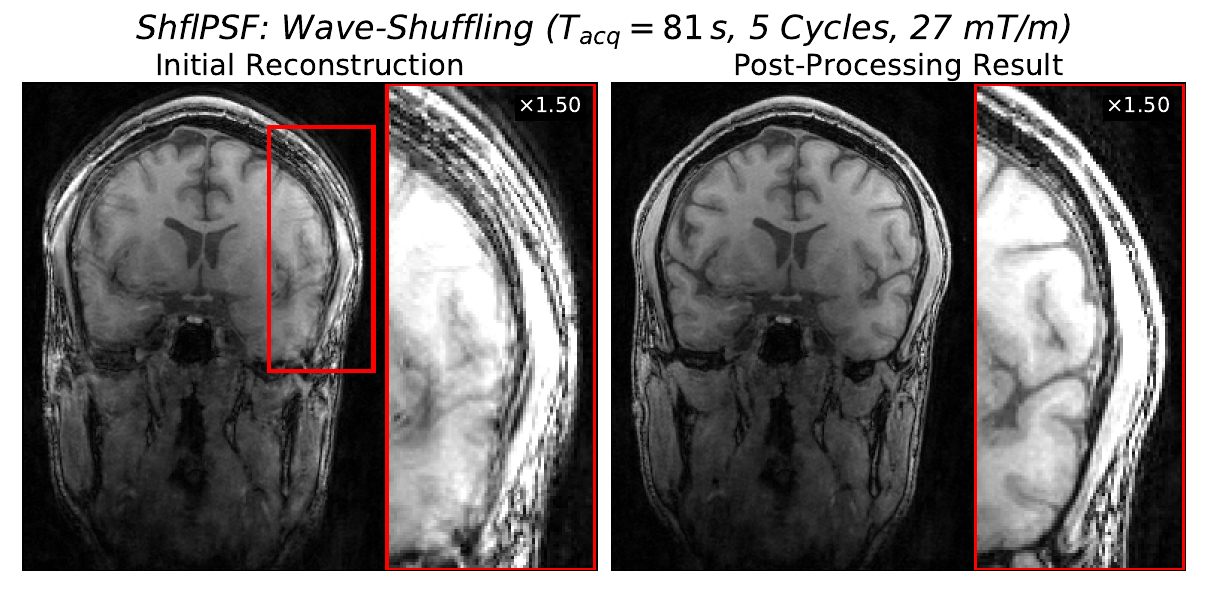}{0.98}{fig:psf_calibration}{
  This figure depicts ShflPSF applied to the 5 cycles, 27 mT/m maximum gradient amplitude MPRAGE
  Wave-Shuffling acquisition at $T_{acq} = 81 s$.
  ShflPSF is able to correct for gradient hardware errors and mitigate ringing artifacts related to
  incorrect wave calibration by performing deconvolution with $W_e^*$ as described in the
  \nameref{sec:autopsf} subsection.
}

\newpage

\fig{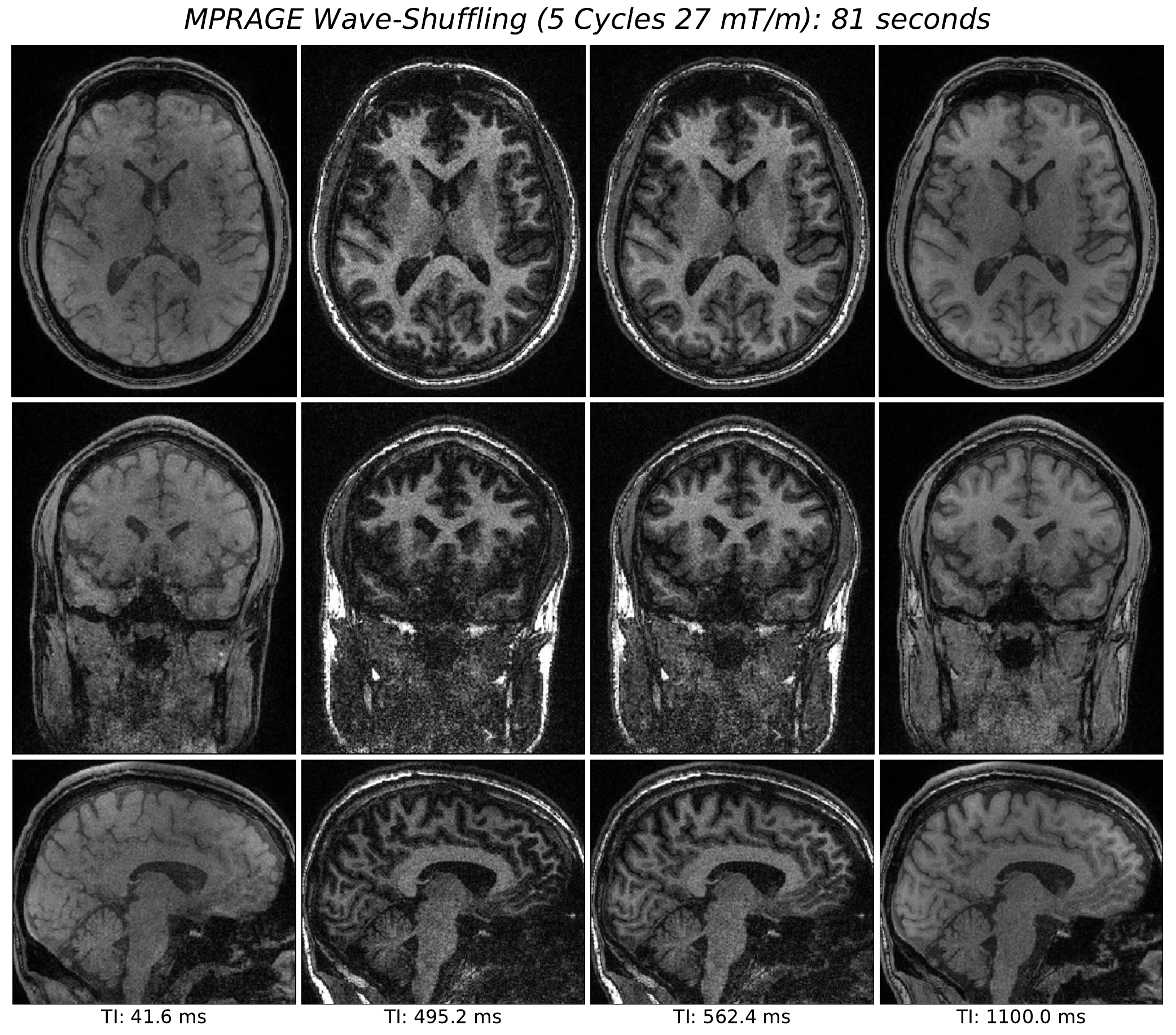}{0.98}{fig:tfl_showcase}{
  Wave-Shuffling achieves 1 mm-isotropic resolution, time-resolved, multi-contrast MPRAGE imaging at
  high acceleration.
  Four out of 256 reconstructed images are depicted.
}

\newpage

\fig{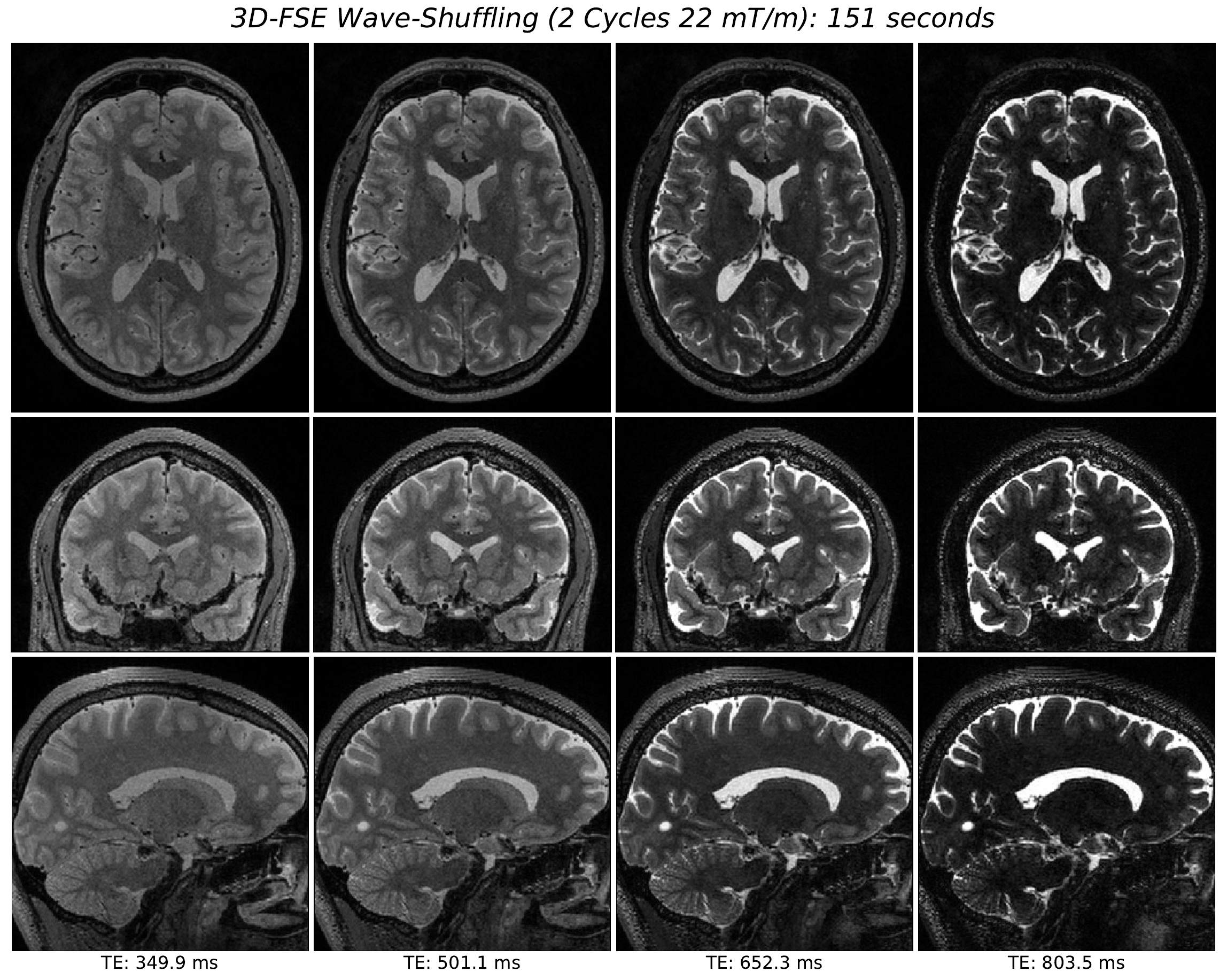}{0.98}{fig:tse_showcase}{
  Wave-Shuffling achieves 1 mm-isotropic resolution, time-resolved, multi-contrast 3D-FSE
  imaging at high acceleration.
  Four out of 256 reconstructed images are depicted.
}

\clearpage

\listoffigures

\end{document}